\documentclass[
reprint,
superscriptaddress,
showkeys,
showpacs,
bibnotes,
amsmath,amssymb,
aps,
prl,
]{revtex4-2}

\usepackage{graphicx}
\usepackage{sidecap}
\usepackage{dcolumn}
\usepackage{upgreek}
\usepackage{mathrsfs}
\usepackage{amsmath}
\usepackage{setspace}%
\usepackage{bm}
\usepackage[breaklinks=true,colorlinks=true,linkcolor=blue,urlcolor=blue,citecolor=blue]{hyperref}
\allowdisplaybreaks[4]

\makeatletter
\renewcommand*\env@matrix[1][\arraystretch]{%
	\edef\arraystretch{#1}%
	\hskip -\arraycolsep
	\let\@ifnextchar\new@ifnextchar
	\array{*\c@MaxMatrixCols c}}
\makeatother

\begin{document}
	\preprint{APS/123-QED}
	
	
	\title{Manipulating synthetic gauge fluxes via multicolor dressing of Rydberg-atom arrays}
	
	
	\author{Xiaoling Wu}
	\affiliation{State Key Laboratory of Low Dimensional Quantum Physics, Department of Physics, Tsinghua University, Beijing 100084, China}

	\author{Fan Yang}
	\thanks{The first two authors contributed equally to this work}
	\email{fanyang@phys.au.dk}
	\affiliation{Center for Complex Quantum Systems, Department of Physics and Astronomy, Aarhus University, DK-8000 Aarhus C, Denmark}

	\author{Shuo Yang}
	\email{shuoyang@mail.tsinghua.edu.cn}
	\affiliation{State Key Laboratory of Low Dimensional Quantum Physics, Department of Physics, Tsinghua University, Beijing 100084, China}
	\affiliation{Frontier Science Center for Quantum Information, Beijing 100084, China}
	\affiliation{Collaborative Innovation Center of Quantum Matter, Beijing 100084, China}
	\affiliation{Hefei National Laboratory, Hefei, 230088, China}
	
	\author{Klaus M{\o}lmer}
	\affiliation{Center for Complex Quantum Systems, Department of Physics and Astronomy, Aarhus University, DK-8000 Aarhus C, Denmark}
	\affiliation{Aarhus Institute of Advanced Studies, Aarhus University, DK-8000 Aarhus C, Denmark}

	\author{Thomas Pohl}
	\affiliation{Center for Complex Quantum Systems, Department of Physics and Astronomy, Aarhus University, DK-8000 Aarhus C, Denmark}
	
	\author{Meng Khoon Tey}
	\affiliation{State Key Laboratory of Low Dimensional Quantum Physics, Department of Physics, Tsinghua University, Beijing 100084, China}
	\affiliation{Frontier Science Center for Quantum Information, Beijing 100084, China}
	\affiliation{Collaborative Innovation Center of Quantum Matter, Beijing 100084, China}
	\affiliation{Hefei National Laboratory, Hefei, 230088, China}

	\author{Li You}
	\email{lyou@mail.tsinghua.edu.cn}
	\affiliation{State Key Laboratory of Low Dimensional Quantum Physics, Department of Physics, Tsinghua University, Beijing 100084, China}
	\affiliation{Frontier Science Center for Quantum Information, Beijing 100084, China}
	\affiliation{Collaborative Innovation Center of Quantum Matter, Beijing 100084, China}
	\affiliation{Hefei National Laboratory, Hefei, 230088, China}	
	\affiliation{Beijing Academy of Quantum Information Sciences, Beijing 100193, China}
	
	\begin{abstract}
		Arrays of highly excited Rydberg atoms can be used as powerful quantum simulation platforms. Here, we introduce an approach that makes it possible to implement fully controllable effective spin interactions in such systems. We show that optical Rydberg-dressing with multicolor laser fields opens up distinct interaction channels that enable complete site-selective control of the induced interactions and favorable scaling with respect to decoherence. We apply this method to generate synthetic gauge fields for Rydberg excitations, where the effective magnetic flux can be manipulated at the single-plaquette level by simply varying the phase of the local dressing field. The system can be mapped to a highly anisotropic Heisenberg model, and the resulting spin interaction opens the door for explorations of topological phenomena with nonlocal density interactions. A remarkable consequence of the interaction is the emergence of topologically protected long-range doublons, which exhibit strongly correlated motion in a chiral and robust manner.
	\end{abstract}
	
	
	
	\maketitle
	
	Quantum simulators make it possible to solve many-body problems that are otherwise intractable by classical calculations \cite{georgescu2014quantum}. An important class of such problems arises from quantum lattice models in external gauge fields, in which a host of exotic phases have been predicted, ranging from fractional quantum Hall (FQH) effects \cite{stormer1999fractional,nayak2008non,zhang1989effective} to chiral spin liquids \cite{wen1989chiral,laughlin1990properties,bauer2014chiral,gong2014emergent}. The technological and fundamental significance of such topological phases of matter has motivated significant efforts towards implementing strong synthetic gauge fields in different physical systems
	\cite{mueller2004artificial,aidelsburger2011experimental,miyake2013realizing,aidelsburger2013realization,stuhl2015visualizing,struck2012tunable,kolovsky2011creating,jotzu2014experimental,goldman2014periodically,roushan2017chiral,ozawa2019topological,fang2012photonic,fang2012realizing,yang2016nonreciprocal,liu2020topological,schmidt2015optomechanical,manovitz2020quantum,mathew2020synthetic}.
	
	Rydberg atoms, held in optical lattices or arrays of optical tweezers, are currently among the most promising and versatile simulation platforms for quantum magnetism
	\cite{schauss2018quantum,wu2020concise,morgado2021quantum,browaeys2020many,scholl2021quantum,bluvstein2021controlling,ebadi2021quantum,bharti2022ultrafast}. Realizing an effective magnetic flux for Rydberg excitations requires a complex-valued exchange interaction between different atomic states. This can be accomplished via a proper tuning of dipolar exchange interactions between different Rydberg states \cite{peter2015topological}. By applying a real magnetic field, the intrinsic spin-orbit coupling of such interactions \cite{lienhard2020realization} can be used to generate an effective spin-lattice model with a nonvanishing Peierls phase that emerges perturbatively to leading order in the strength of the dipole-dipole interaction. The possibility to engineer synthetic gauge fields in interacting spin lattices then holds exciting perspectives for exploring exotic many-body phases and dynamics \cite{weber2018topologically,weber2021topological,ohler2022self,weber2022experimentally,ohler2022quantum}.
	\begin{figure}[b!]
		\centering
		\includegraphics[width=\linewidth]{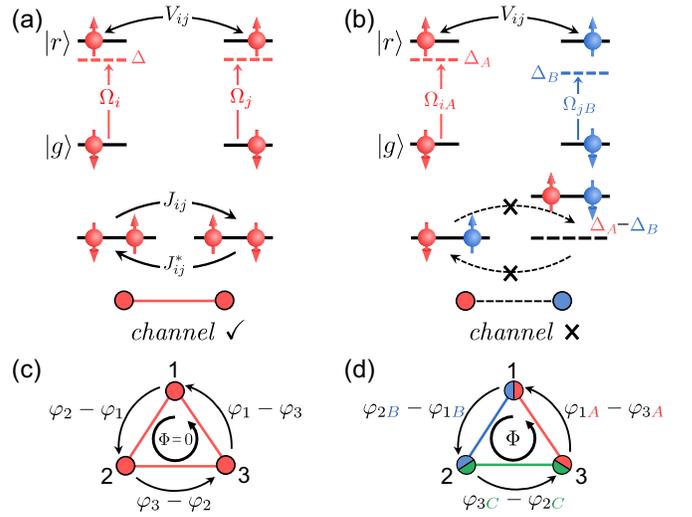}
		\caption{(a) An excitation hopping channel between atoms can be established with monochromatic Rydberg dressing. (b) Lasers with different colors (frequencies) $A$ and $B$ cannot form a hopping channel due to the large energy defect $|\delta E|=|\Delta_{A}-\Delta_{B}|$. (c) Cancellation of the Peierls phase leads to a vanishing magnetic flux $\Phi=0$ in monochromatic dressing. (d) A tunable flux $\Phi$ is realized by the multicolor dressing. Two-colored circles represent simultaneous dressing of the atoms by both laser fields of the colors indicated.}
		\label{fig:fig_1}
	\end{figure}
	
	Here, we describe an approach to implementing synthetic gauge fields via photon-assisted excitation exchange in Rydberg-atom arrays. Dressing the atoms by multicolor laser fields is shown to offer complete  optical control of the generated spin-exchange interactions at the level of individual sites. The induced Peierls phase is determined by the relative phase between the applied laser fields, with which the effective gauge flux can be tuned to arbitrary patterns. Together with the nonlocal density interaction between Rydberg excitations, this yields a versatile quantum simulation approach for exploring strongly correlated systems with nontrivial band topologies and finite-range interactions. We illustrate these perspectives by discussing the topologically protected chiral motion of bound pairs \cite{salerno2018topological,salerno2020interaction,berti2021topological} that emerge from the multicolor Rydberg dressing of an atomic square lattice \cite{aidelsburger2015measuring}.
	\begin{figure}
		\centering
		\includegraphics[width=\linewidth]{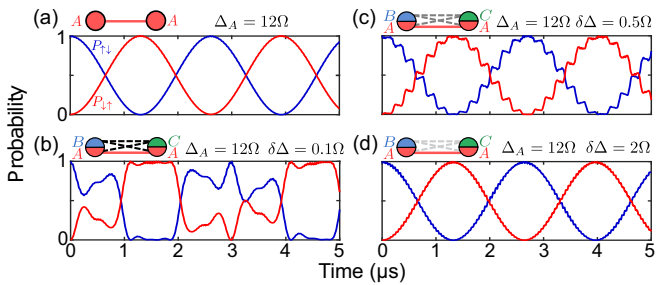}
		\caption{Time evolution of excitation transport for monochromatic dressing in (a), and for multicolor dressing in (b)-(d) with increasing frequency separations $\delta\Delta=\Delta_B-\Delta_A=\Delta_C-\Delta_B$ between colors at induced hopping strength $|J|\approx 0.2\Omega$. }
		\label{fig:fig_2}
	\end{figure}
	
	We consider a two-dimensional array of individually trapped atoms. For a monochromatic Rydberg dressing \cite{zeiher2016many}, the ground state $|g_{i}\rangle$ of the $i$th atom (located at $\mathbf{r}_{i}$) is coupled to a Rydberg state $|r_{i}\rangle$ by an off-resonant laser field with Rabi frequency $\Omega_{i
	}=|\Omega_{i}|e^{i\varphi_{i}}$ and detuning $\Delta=\omega_{0}-\omega_{L}$, where $\omega_{0}$ and $\omega_{L}$ are the atomic transition frequency and laser frequency, respectively. Let us first consider a two-atom model [see Fig.~\ref{fig:fig_1} (a)]. In this case, the singly excited pair states $|g_i r_j\rangle$ and $|r_i g_j\rangle$ are degenerate and coupled by two Raman pathways with intermediate states $|g_i g_j\rangle$ and $|r_i r_j\rangle$ \cite{yang2019quantum,PhysRevLett.125.143601}. Importantly, the contribution from both pathways is not symmetric because the doubly excited state $|r_i r_j\rangle$ is shifted by the van der Waals (vdW) interaction $V_{ij}$ between Rydberg atoms. In the limit of large detunings, $|\Omega_{i(j)}|\ll |\Delta|, |\Delta + V_{ij}|$, this asymmetry leads to an effective hopping of the Rydberg excitation from site $j$ to site $i$ with a strength $|J_{ij}|= |\Omega_i\Omega_j^* V_{ij}/ 4\Delta(\Delta+V_{ij})|$. Taking into account the phase of each dressing field, this hopping amplitude $J_{ij}=|J_{ij}|e^{i\phi_{ij}}$ becomes complex, with a Peierls phase $\phi_{ij}=\varphi_{i}-\varphi_{j}$ given by the sum of the phases $\varphi_{i}$ and -$\varphi_{j}$ of both involved laser fields. However, it is impossible to induce a nonvanishing effective magnetic flux $\Phi$ because the same driving field creates and annihilates an excitation on each site, leading to a vanishing net flux as the excitation circulates around a closed loop. This is illustrated in Fig.~\ref{fig:fig_1}(c) for three atoms whose total flux $\Phi=(\varphi_{2}-\varphi_{1})+(\varphi_{3}-\varphi_{2})+(\varphi_{1}-\varphi_{3})=0$ vanishes identically. 
	
	As we shall show below, a finite magnetic flux can, however, be realized by applying multiple dressing fields with different frequencies (colors). The underlying mechanism exploits the frequency sensitivity of the stimulated excitation exchange, which is only resonant if the two involved atoms share a set of dressing fields with identical colors. As illustrated in Fig.~\ref{fig:fig_1}(d) for a three-atom plaquette, this makes it possible to independently control the phase of the exchange interaction between each atom pair with only three different colors $A$, $B$, $C$ of the dressing fields applied, i.e., driving each atom with two different colors opens up a distinct interaction \emph{channel} for each atom pair. The phase of each laser field can be tuned individually, such that we obtain a finite and tunable flux $\Phi=(\varphi_{2B}-\varphi_{1B})+(\varphi_{3C}-\varphi_{2C})+(\varphi_{1A}-\varphi_{3A})$. Here, we have labeled the dressing fields of different colors by $\Theta \in \{A,B,\cdots\}$ and denote their phases at a given site $i$ by $\varphi_{i\Theta}$.

\begin{figure}
		\centering
		\includegraphics[width=\linewidth]{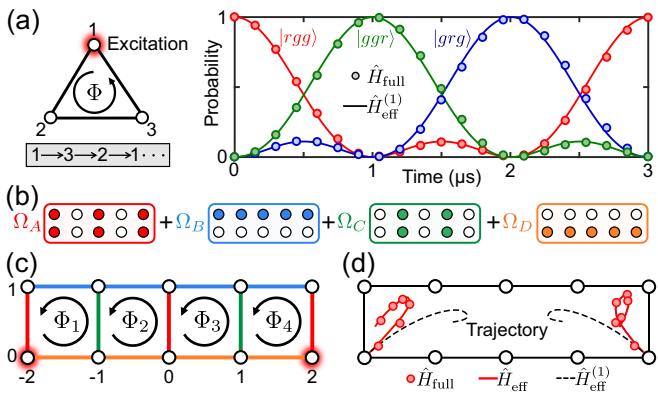}
		\caption{(a) Single-excitation chiral motion in a three-atom plaquette at $\Phi=\pi/2$, with the chirality indicated by the grey box. (b) Dressing schemes to implement a Harper-Hofstadter lattice with an arbitrary flux pattern shown in (c). (d) Center-of-mass trajectories for an excitation in the left (${x}\leq 0$) and right (${x}\geq 0$) regions at $\Phi_1=\Phi_4=\pi/3$, $\Phi_2=\Phi_3=\pi/2$. The detunings are $(\Delta_{A},\Delta_{B},\Delta_{C},\Delta_{D}) = 2\pi \times (120, 140, 160, 180)~\mathrm{MHz}$ in (a) and (d), with Rabi frequencies $(\Omega_{A},\Omega_{B},\Omega_{C}) = 2\pi \times (10, 10.9, 11.7)~\mathrm{MHz}$ in (a) and $(\Omega_{A},\Omega_{B},\Omega_{C},\Omega_{D}) = 2\pi \times (10, 11.5, 13.1, 14.6)~\mathrm{MHz}$ in (d) \cite{para}.}
		\label{fig:fig_3}
	\end{figure}

	We can gain more intuition by considering two atoms that are each dressed by fields with distinct detunings $\Delta_A$ and $\Delta_B$ [see Fig.~\ref{fig:fig_1} (b)]. This implies a finite energy defect $\delta E=\Delta_{A}-\Delta_{B}$ for the excitation transfer between the states $|g_{i} r_{j}\rangle$ and $|r_{i} g_{j}\rangle$. The stimulated emission and reabsorption of photons is, thus, off-resonant by $|\delta E|$ and will suppress the excitation hopping if $|\delta E|\gg |J_{ij}|$. Under this condition we can introduce distinct interaction channels that only emerge when a pair of atoms is dressed by laser-field with the same color  $\Theta^{[ij]}$ \cite{Supplement}, e.g., $\Theta^{[23]}=\{B,C\} \cap \{C,A\}=C$ for the configuration considered in Fig.~\ref{fig:fig_1}(d). To verify that the crosstalk between different colors can indeed be suppressed, we calculate the dynamics governed by the exact Hamiltonian
	\begin{equation}
	\hat{H}_{\mathrm{full}}(t)=\sum_{i}\hat{H}_{i}(t)+\sum_{i<j}V_{ij}|r_{i}\rangle\langle r_{i}|\otimes |r_{j}\rangle\langle r_{j}|
	\label{eq:eq1}
	\end{equation}
	for two dressed atoms. Here,
	\begin{align}
	\hat{H}_{i}(t)&=\sum_{\Theta}\left(\frac{\Omega_{i\Theta
	}}{2}e^{i\Delta_{\Theta}t}|r_{i}\rangle\langle g_{i}|+\mathrm{H.c}\right) \label{eq:eq2}
	\end{align} 
	describes the time-dependent single-particle Hamiltonian of the $i$th atom, and $V_{ij}={C_{6}}/{|\mathbf{r}_{i}-\mathbf{r}_{j}|^6}$ denotes the vdW interaction between Rydberg excitations. As shown in Fig.~\ref{fig:fig_2}(a), the excitation transport is nearly perfect for monochromatic dressing. The inclusion of two additional fields with different colors $B$ and $C$, does not perturb this transport as long as the corresponding energy offset $|\Delta_{\Theta}-\Delta_{\Theta^{\prime}}|$ is much larger than the respective hopping strength [Figs.~\ref{fig:fig_2}(b)-\ref{fig:fig_2}(d)]. With feasible experimental parameters \cite{para}, this condition can be well fulfilled, while realizing a flux of $\Phi=\pi/2$ for the three-atom case considered in Fig.~\ref{fig:fig_1}(d). This is verified by Fig.~\ref{fig:fig_3}(a), where we find a characteristic chiral motion of the Rydberg excitation around the plaquette based on exact simulation of the dynamics. The simulation is in good agreement with an effective model that neglects any crosstalk between different colors.
	
	We can formulate the corresponding effective Hamiltonian by a perturbative analysis using Floquet theory \cite{bukov2015universal,eckardt2015high,rudner2020floquet,li2021coherent,goldman2014periodically,son2009floquet,ho1985semiclassical,aravind1984two,shavitt1980quasidegenerate}. Introducing bosonic creation operators $\hat{b}_{i}^\dagger=|r_{i}\rangle \langle g_{i}|$ with the hard-core constraint $\hat{b}_i^\dagger\hat{b}_i^\dagger=0$, we obtain a single-excitation effective Hamiltonian
	\begin{equation}
	\hat{H}_{\mathrm{eff}}^{(1)}=\sum_{i}\mu_{i}\hat{b}_{i}^{\dagger}\hat{b}_{i}+
	\sum_{i<j,\Theta^{[ij]}}(e^{i\phi_{ij\Theta^{[ij]}}}|J_{ij\Theta^{[ij]}}|\hat{b}_{i}^{\dagger}\hat{b}_{j}+\mathrm{H.c.}),\label{eq:eq4}
	\end{equation}
	where $\phi_{ij\Theta^{[ij]}}=\varphi_{i\Theta^{[ij]}}-\varphi_{j\Theta^{[ij]}}$ denotes the Peierls phase,
	\begin{equation}
	|J_{ij\Theta^{[ij]}}|=\left|\frac{\Omega_{i\Theta^{[ij]}}\Omega^*_{j\Theta^{[ij]}} V_{ij}}{4\Delta_{\Theta^{[ij]}}(\Delta_{\Theta^{[ij]}}+V_{ij})}\right|\label{eq:eq6}
	\end{equation}
	is the hopping strength of the excitation, and $\mu_i$ is the on-site chemical potential. All these parameters can be tuned individually by adjusting the dressing fields at each site \cite{Supplement}. Importantly, the effective magnetic flux can be tuned by controlling the phase distribution of the applied dressing fields and does not depend on the laser intensities, lattice configurations, or Rydberg interactions. This makes it possible to generate an arbitrary magnetic flux pattern while implementing the effective Hamiltonian to an arbitrary accuracy by independently suppressing state leakages and crosstalk errors. Furthermore, the U(1) symmetry of the effective Hamiltonian Eq.~(\ref{eq:eq4}) protects the system against various decoherence processes, e.g., global laser phase noise induced dephasing is eliminated in the resulting decoherence-free subspace \cite{PhysRevLett.81.2594,Supplement,PhysRevA.105.062602}, and the damping caused by decay of the Rydberg states can be mitigated by post-selection as described in the Supplemental Material \cite{Supplement}. In the Supplemental Material \cite{Supplement}, we show that the challenging single-site control of the laser intensity and phase distribution can be achieved by a compact optical module, whose complexity does not grow with system size. We also provide explicit schemes for manipulating gauge fluxes in different lattice geometries, including a square lattice, a triangular lattice, and a honeycomb lattice \cite{Supplement}. Altogether, this offers an accurate and scalable approach to synthesizing gauge fields with promising coherence properties.

	\begin{figure*}
		\centering
		\includegraphics[width=0.98\linewidth]{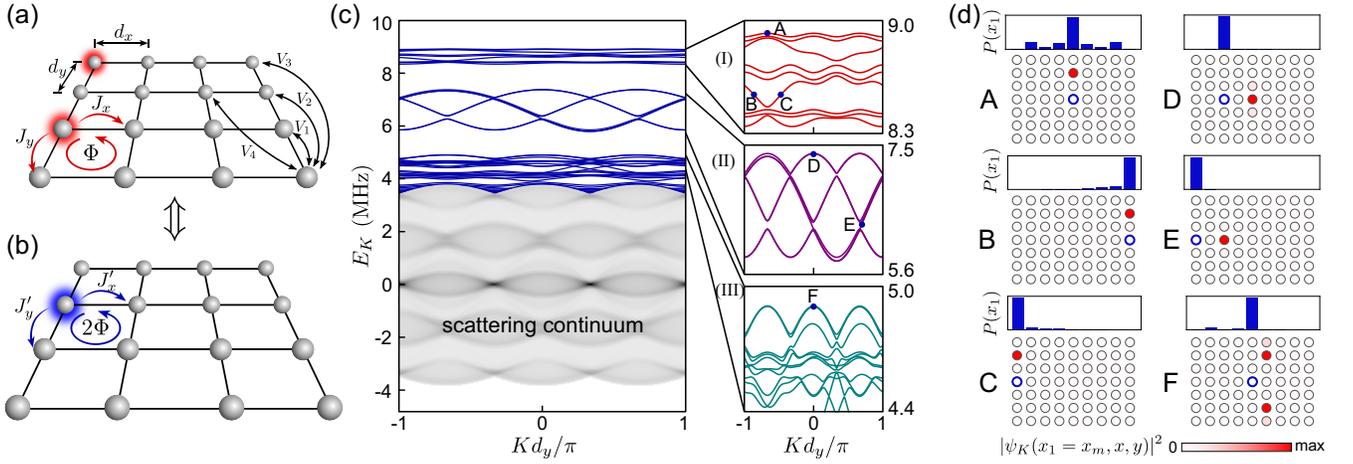}
		\caption{(a) Anisotropic Harper-Hofstadter model with lattice constants $d_x$ and $d_y$ along $x$ and $y$ directions. A feasible implementation with $d_x=5.1~\mu\mathrm{m}$, $d_y=4.8~\mu\mathrm{m}$, Rabi frequencies $\sim 2\pi\times 10~$MHz, and a $^{87}$Rb-atom Rydberg state $|70S_{1/2}\rangle$ yields typical parameters $J_x=0.5~$MHz, $J_x/J_y=0.6$, and $(V_1,V_2,V_3,V_4)/J_x=(1045,16,1.4,7.7)$. (b) Effective lattice for the center-of-mass motion of the bound pair shown in (a). (c) Two-body spectra of the system with 9 sites along the $x$ direction, where different types (I)-(III) of bound pairs are distinguished by their orientations. (d) Density distribution for the first excitation [$P(x_1)=\sum_{x,y}|\psi_K(x_1,x,y)|^2$] and the second excitation ($|\psi_K(x_1=x_m,x,y)|^2$, scaled by shades of red) with $x_m=\arg_\mathrm{max}[P(x_1)]$ the most occupied site of the first excitation (blue empty circle).}
		\label{fig:fig_y1}
	\end{figure*}
	Let us now discuss the behavior of multiple excitations, which feature strong vdW interactions between them.
	If the initial distance between excitations is sufficiently large such that their interactions $V_{ij}$ are smaller than the detunings \cite{Supplement}, the simple Hamiltonian 
	\begin{equation}
	\hat{H}_{\mathrm{eff}}=\hat{H}_{\mathrm{eff}}^{(1)}+\sum_{i<j}V_{ij}\hat{b}_{i}^{\dagger}\hat{b}_{j}^{\dagger}\hat{b}_{j}\hat{b}_{i} \label{eq:Heff}
	\end{equation}
	describes the many-body dynamics of the system. Such a Hamiltonian is equivalent to a highly anisotropic Heisenberg model, as the density interaction $V_{ij}$ is two orders of magnitude larger than the hopping strength $|J_{ij}|$. The nonlocal density interaction $V_{ij}$ also facilitates the study of many-body dynamics beyond the hard-core constraints. To verify the effective model and its distinct features, we first consider to implement a Harper-Hofstadter ladder. With dressing fields of four different colors introduced in Fig.~\ref{fig:fig_3}(b), an arbitrary flux pattern can be realized [Fig.~\ref{fig:fig_3}(c)]. For a symmetric pattern $\Phi_1=\Phi_4$, $\Phi_2=\Phi_3$, we calculate the trajectories for two colliding excitations initially localized at the corners. The exact simulation based on Eq.~(\ref{eq:eq1}) shows significant repulsion between the excitations [Fig.~\ref{fig:fig_3}(d)], in agreement with the calculation based on Eq.~(\ref{eq:Heff}). We also note that the trajectories for two excitations are no longer symmetric when nonlocal interaction $V_{ij}$ comes into play \cite{yu2017symmetry,tai2017microscopy}. Such an interesting feature does not appear in systems with only hard-core constraints [Eq.~(\ref{eq:eq4})].
	

	The ability to implement such effective models permits exploration of topological phenomena in the presence of strong and finite-range interactions. As an example [Fig.~\ref{fig:fig_y1}(a)], we consider two-excitation dynamics in an anisotropic Harper-Hofstadter lattice \cite{liang2021coherence}, where the nearest-neighbor hopping take on different strengths $J_x$ and $J_y$ along $x$ and $y$ directions, respectively. The model can be realized with the similar scheme shown in Fig.~\ref{fig:fig_3}(b).
	When the flux $\Phi$ is uniform and rational ($\Phi/2\pi=p/q$), the single-particle spectra split into $q$ gapped topological bands linked by gapless edge modes. Remarkably, the finite-range density interaction can further lead to the emergence of chiral edge-mode bound states with large bond lengths, as we will show below.
	
	Let us consider first the case of two tightly bound excitations, separated by two lattice sites along the $y$ direction as illustrated in Fig.~\ref{fig:fig_y1}(a). When the differences $D_1=V_{2}-V_{1}$, $D_2=V_{2}-V_{4}$, and $D_3 = V_{2}-V_{3}$ between density interactions $V_2$ and $V_1$, $V_3$, $V_4$ [Fig.~\ref{fig:fig_y1}(a)] are much larger than the tunneling strength, we can perturbatively construct an effective Harper-Hofstadter lattice for the center-of-mass dynamics of the bound pair [Fig.~\ref{fig:fig_y1}(b)], which moves across the lattice with modified hopping strengths $J_x^\prime=2J_x^2/D_2$ and $J_y^\prime=J_y^2/D_1+J_y^2/D_3$. Most importantly, the effective magnetic flux $\Phi^\prime = 2\Phi$ of the doublon increases to twice the single-particle flux. An interesting case emerges for  $\Phi=2\pi/3$ such that $\Phi^\prime=4\pi/3~(\mathrm{mod}~2\pi)=-\Phi$. In this case we expect to observe the co-existence of single-excitation chiral edge state and topologically protected edge-mode bound state with opposite chirality.
	
	When the above interaction difference is finite, the situation becomes complicated and an exact diagonalization is required \cite{letscher2018mobile,macri2021bound}. Since the effective Hamiltonian is verified in Fig.~\ref{fig:fig_3}(d), we will perform the calculation based on Eq.~(\ref{eq:Heff}). For an infinitely extended lattice along the $y$ direction with a finite number of sites along the $x$ direction, the two-body eigenstate with a $y$-direction center-of-mass quasimomentum $K$ can be written as
	\begin{equation}
	|\psi_K\rangle=\frac{1}{{2\pi}}\sum_{\mathbf{r}_1\neq\mathbf{r}_2}\psi_K(x_1,x_2,r)e^{iKR}\hat{b}_{\mathbf{r}_1}^\dagger\hat{b}_{\mathbf{r}_2}^\dagger|0\rangle,
	\end{equation}
	where $r=y_2-y_1$ and $R=(y_1+y_2)/2$ denote the relative and center-of-mass coordinates along the $y$ direction, $\hat{b}_{\mathbf{r}_\nu}^\dagger$ creates a Rydberg excitation at $\mathbf{r}_\nu=(x_\nu,y_\nu)$, and $|0\rangle$ is the vacuum state with all atoms in $|g\rangle$. The associated spectrum of eigenenergies $E_K$ is shown in Fig.~\ref{fig:fig_y1}(c), and reveals a number of interesting states. The scattering continuum forms the lowest band with oscillating density of states. Above the scattering continuum, we identify different patterns of bound states. Figures~\ref{fig:fig_y1}(c) and \ref{fig:fig_y1}(d) show the dispersion and density profile of three types of these states. The type-I bound state corresponds to the one depicted in Fig.~\ref{fig:fig_y1}(a), whose spectrum is split into three energy bands with Chern numbers $\mathcal{C}=\{-1,2,-1\}$ (from the lowest to the highest) predicted by the center-of-mass motion analysis.
	
	The state, marked as A in Fig.~\ref{fig:fig_y1}(c,I), is located in the upper band and represents a normal bound state in the bulk of the system. The other two states, marked as B and C, lie within the gap between the lowest and the middle bands and are topologically protected bound states that are respectively localized at the right and left edge of the lattice. Similar states can be identified for the type-II bound pair [Fig.~\ref{fig:fig_y1}(c,II)], which are aligned along the $x$ direction, and form bulk-mode (marked as D) as well as chiral-edge-mode (marked as E) bound states. The finite range of the Rydberg-state interaction also makes it possible to form bound pairs separated by a longer distance, such as type-III state, indicated in Fig.~\ref{fig:fig_y1}(c,III) and shown in panel F of Fig.~\ref{fig:fig_y1}(d).

\begin{figure}
		\centering
		\includegraphics[width=\linewidth]{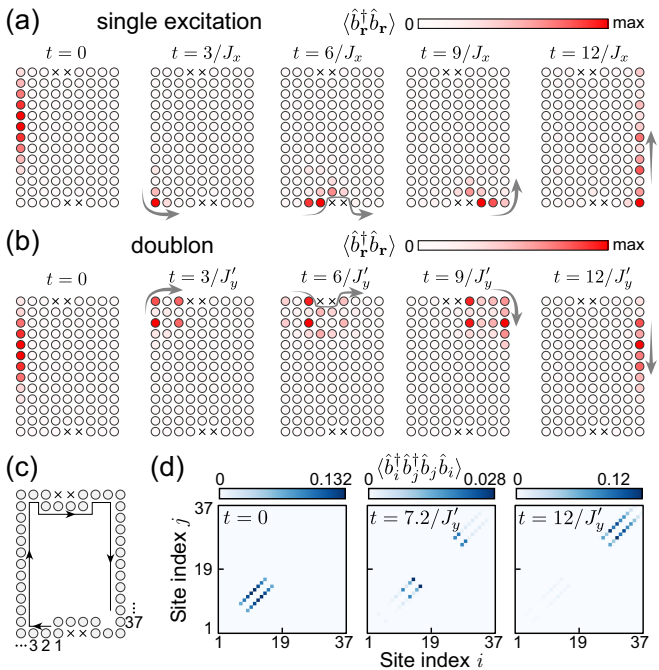}
		\caption{(a) and (b) show evolution of the density distribution $\langle \hat{b}_\mathbf{r}^\dagger \hat{b}_\mathbf{r}\rangle$ for the lowest chiral edge mode of a single-excitation and a type-I bound pair, respectively. ``$\times$'' denotes a vacancy defect. (c) Definition of the site index along the edge. (d) Evolution of the two-body correlation function $\langle \hat{b}_{i}^\dagger \hat{b}_{j}^\dagger \hat{b}_{j} \hat{b}_{i}\rangle$.}
		\label{fig:fig_y2}
\end{figure}

	Fig.~\ref{fig:fig_y2}(a) shows the transport dynamics of a single-excitation edge mode and demonstrates counter-clock wise propagation. The depicted motion around a finite lattice with edge defects indeed shows negligible backward scattering. The dynamics of the type-I bound state C [Figs.~\ref{fig:fig_y1}(c,I) and \ref{fig:fig_y1}(d)] is displayed in Fig.~\ref{fig:fig_y2}(b). One clearly finds unidirectional doublon motion that is robust against local defects, and now features opposite chirality compared to the single-excitation transport. Fig.~\ref{fig:fig_y2}(d) shows the two-body correlation function along the outer edge of the lattice [Fig.~\ref{fig:fig_y2}(c)], and demonstrates that the bound-state structure of the doublon indeed stays fully intact during its topologically protected transport.
	
	In conclusion, we have described the application of multicolor laser-dressing to generate synthetic gauge fields for Rydberg excitations in neutral atom quantum simulators. Our approach makes it possible to realize arbitrary Peierls phases without compromising the accuracy of the quantum simulation. In particular, the scheme features individual tunability of the magnetic flux, which permits the experimental study of rich physics not yet explored with a non-uniform gauge field, e.g., exotic metal-insulator transition driven by the random flux \cite{zhang1994effective,furusaki1999anderson,li2022transition}, and emergence of anyons from a fractal flux pattern \cite{manna2022anyon}. The inherent long-range interacting feature also motives future work on topological phenomena in interacting systems, such as emergent dynamical gauge fields \cite{ohler2022self,Lukas2022quantum} and fractional quantum Hall physics \cite{yao2012topolo}.
	
	{\it Note added}. We recently became aware of two interesting schemes for realizing synthetic gauge fields in Rydberg-atom arrays, respectively based on laser-assisted dipole-dipole interactions in a rectangular lattice \cite{yang2022quantum} and Rydberg-dressing induced ground-state interactions in a honeycomb lattice \cite{zhao2022fractional}.
	
	\begin{acknowledgments}
		We thank Xiangliang Li, Xinhui Liang, Cheng Chen, Anne E. B. Nielsen, Anton L. Andersen, Jinlong Yu, Shuaifeng Guo, Feng Chen, Xin-Chi Zhou, and Rui Lin for valuable discussions. This work is supported by the National Key R$\&$D Program of China (Grant No.~2018YFA0306504 and No.~2018YFA0306503), the National Natural Science Foundation of China (NSFC) (Grant No. 12174214), and by the Innovation Program for Quantum Science and Technology (Project 2-9-4). F. Yang, K. M{\o}lmer, and T. Pohl acknowledge the support from Carlsberg Foundation through the ``Semper Ardens'' Research Project QCooL and from the Danish National Research Foundation (DNRF) through the Center of Excellence ``CCQ'' (Grant No.~DNRF156).
	\end{acknowledgments}

	\bibliography{ref}
\end{document}


\preprint{APS/123-QED}
	
	\title{Supplemental Material for ``Manipulating synthetic gauge fluxes via multicolor dressing of Rydberg-atom arrays''}
	
	\author{Xiaoling Wu}
	\thanks{These authors contributed equally to this work.}
	\affiliation{State Key Laboratory of Low Dimensional Quantum Physics, Department of Physics, Tsinghua University, Beijing 100084, China}

	\author{Fan Yang}
	\thanks{These authors contributed equally to this work.}
	\affiliation{Center for Complex Quantum Systems, Department of Physics and Astronomy, Aarhus University, DK-8000 Aarhus C, Denmark}

	\author{Shuo Yang}
	\affiliation{State Key Laboratory of Low Dimensional Quantum Physics, Department of Physics, Tsinghua University, Beijing 100084, China}
	\affiliation{Frontier Science Center for Quantum Information, Beijing 100084, China}
	\affiliation{Collaborative Innovation Center of Quantum Matter, Beijing 100084, China}
	\affiliation{Hefei National Laboratory, Hefei, 230088, China}	
	
	\author{Klaus M{\o}lmer}
	\affiliation{Center for Complex Quantum Systems, Department of Physics and Astronomy, Aarhus University, DK-8000 Aarhus C, Denmark}
	\affiliation{Aarhus Institute of Advanced Studies, Aarhus University, DK-8000 Aarhus C, Denmark}

	\author{Thomas Pohl}
	\affiliation{Center for Complex Quantum Systems, Department of Physics and Astronomy, Aarhus University, DK-8000 Aarhus C, Denmark}
	
	\author{Meng Khoon Tey}
	\affiliation{State Key Laboratory of Low Dimensional Quantum Physics, Department of Physics, Tsinghua University, Beijing 100084, China}
	\affiliation{Frontier Science Center for Quantum Information, Beijing 100084, China}
	\affiliation{Collaborative Innovation Center of Quantum Matter, Beijing 100084, China}
	\affiliation{Hefei National Laboratory, Hefei, 230088, China}

	\author{Li You}
	\affiliation{State Key Laboratory of Low Dimensional Quantum Physics, Department of Physics, Tsinghua University, Beijing 100084, China}
	\affiliation{Frontier Science Center for Quantum Information, Beijing 100084, China}
	\affiliation{Collaborative Innovation Center of Quantum Matter, Beijing 100084, China}
	\affiliation{Hefei National Laboratory, Hefei, 230088, China}	
	\affiliation{Beijing Academy of Quantum Information Sciences, Beijing 100193, China}
	
	\maketitle
	\onecolumngrid
	
	This Supplemental Material provides additional information for the main text. In Sec.~\ref{sec:sec1}, detailed derivations of the effective Hamiltonians in different regimes are given by using quantum Floquet theory and perturbation analysis. In Sec.~\ref{sec:sec2}, we illustrate how to adjust chemical potentials and the non-nearest neighbour hoppings. In Sec.~\ref{sec:decoherence}, we systematically study the decoherence of our scheme. In Sec.~\ref{sec:sec3}, a feasible experimental implementation is designed for simulating the Harper-Hofstadter model and we give some considerations about technical challenges in the future.
	
	\section{Effective Hamiltonian}\label{sec:sec1}
	In this section, we provide a Floquet analysis \cite{eckardt2015high,rudner2020floquet,goldman2014periodically,son2009floquet} of our system. By using the generalized van Vleck (GVV) nearly degenerate perturbation theory \cite{ho1985semiclassical,aravind1984two,shavitt1980quasidegenerate} to the infinite-dimensional Floquet matrix, time-independent effective Hamiltonians in different regimes are derived. 
	
	\subsection{Floquet formulation}
\begin{figure}
		\centering
		\includegraphics[width=0.6\linewidth]{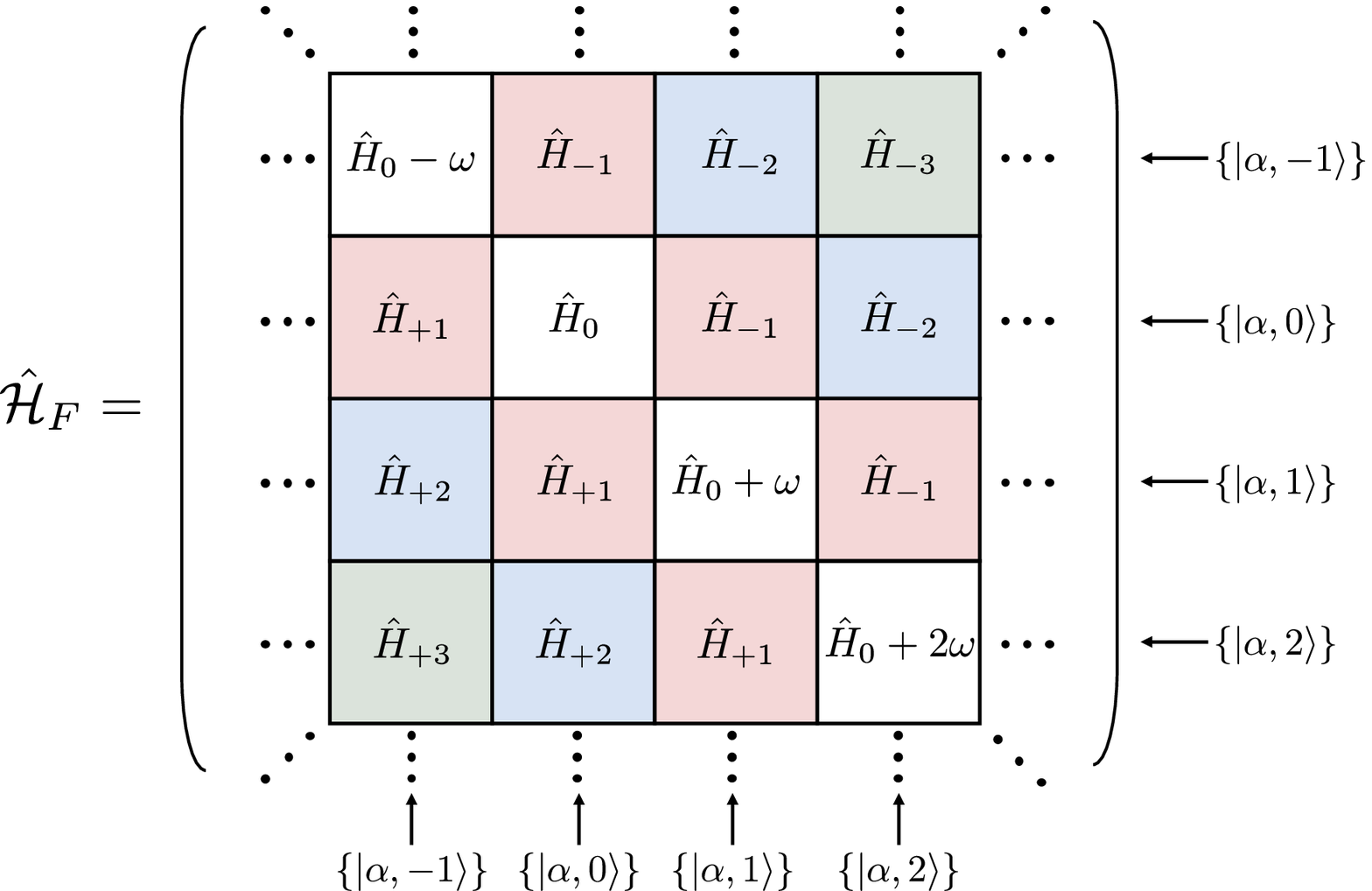}
		\caption{Block structure of the Floquet matrix $\hat{\mathcal{H}}_{F}$. Each block consists of a $L \times L$ matrix, with $L$ the dimension of the physical system. States in diagonal blocks of different Fourier index $n$, such as $|\alpha,n \rangle$ and $|\alpha^\prime,n^\prime \rangle$, are coupled by matrix elements in nondiagonal blocks $\hat{H}_{\pm (n-n^\prime)}$. }
		\label{fig:fig_s1}
	\end{figure}

	Quantum Floquet theory can be applied to quantum systems governed by time-periodic Hamiltonian $\hat{H}(t) = \hat{H}(t+T)$, 
	that satisfies time-dependent Sch{\"o}rdinger equation
	\begin{equation}
		i\hbar\frac{\partial}{\partial t}\psi(t) = \hat{H}(t)\psi(t),\label{eq:eq1}
	\end{equation}
	which possesses a solution in form of $\psi(t) = e^{-i\varepsilon t/\hbar}\phi(t)$, where the so-called \emph{Floquet mode} $\phi(t)$ exhibits the same time periodicity as $\hat{H}(t)$ and $\varepsilon$ is the corresponding quasienergy. The eigenvalue problem for quasienergy is given by
	\begin{equation}
		\left(H(t)-i\hbar\frac{\partial}{\partial t}\right)\phi(t) = \varepsilon\phi(t), \label{eq:eq2}
	\end{equation}
	which can be transformed into an equivalent time-independent problem, with the help of an infinite-dimensional Floquet matrix in an extended Hilbert space \cite{eckardt2015high,son2009floquet}. Following the standard approach, we expand the Hamiltonian $\hat{H}(t)$ and quasienergy eigenfuction $\phi(t)$ into the Fourier components of $\omega=2\pi/T$, 
	\begin{equation}
		\hat{H}(t)=\sum_{n}\hat{H}_{n}e^{-in\omega t},\qquad \qquad   \phi(t)=\sum_{n}\phi_{n}e^{-in\omega t} .\label{eq:eq3}
	\end{equation}
	The extended Hilbert space is then spanned by a complete set of orthonormal basis $	|\alpha, n\rangle = |\alpha\rangle\otimes|n\rangle $, 
	where $|\alpha\rangle$ is a complete set of orthonormal basis states of the original physical system and $n$ denotes the Fourier index ranging from $-\infty$ to $\infty$. The matrix element of the time-independent Floquet Hamiltonian $\hat{\mathcal{H}}_{F}$ can be calculated as
	\begin{equation}
		\langle\alpha^{\prime},n^{\prime}|\hat{\mathcal{H}}_{F}|\alpha, n\rangle=\langle\alpha^{\prime}|\hat{H}_{n^{\prime}-n}|\alpha\rangle + \delta_{n^{\prime}n}\delta_{\alpha^{\prime}\alpha}n\hbar \omega. \label{eq:eq4}
	\end{equation}
	where $\hbar=1$ is assumed unless otherwise noted. The block structure of the Floquet matrix $\hat{\mathcal{H}}_{F}$ is shown in Fig.~\ref{fig:fig_s1}, with each block of the same dimension as the physical system, and a physical state belonging to index $n$ can be coupled to another state belonging to index $n^{\prime}$ by matrix elements of blocks $\hat{H}_{\pm (n^{\prime}-n)}$.
	
	\subsection{The generalized van Vleck nearly degenerate perturbation theory}
	If the Floquet matrix can be divided into an unperturbed part $\hat{\mathcal{H}}_{0}$ and a perturbation $\lambda\hat{\mathcal{V}}$, i.e. $\hat{\mathcal{H}}_{F}=\hat{\mathcal{H}}_{0}+\lambda\hat{\mathcal{V}},$
	the corresponding effective Hamiltonian and its eigenstate solution in the concerned subspace can be expanded in powers of parameter $\lambda$,
	\begin{equation}
		\hat{\mathcal{H}}_{\mathrm{eff}}=\sum_{m=0}^{\infty}\lambda^{m}\hat{\mathcal{H}}_{\mathrm{eff}}^{(m)},\qquad\qquad \Psi = \sum_{m=0}^{\infty}\lambda^{m}\Psi^{(m)}. \label{eq:eq5}
	\end{equation}
	Without loss of generality, we set $\lambda=1$ in the following analysis.
	
	For our system, $\hat{\mathcal{H}}_{0}=\hat{\mathcal{H}}_{\mathrm{eff}}^{(0)}$ includes all diagonal terms derived from the transition energy of Rydberg state as well as interactions between Rydberg atoms. Its eigenstates $\Psi^{(0)}$ form the orthonormal basis $|\alpha,n\rangle$ described above, which can also be considered as $0$th-order eigenstates. The perturbation $\hat{\mathcal{V}}$ represents the remaining nondiagonal coupling terms from driving fields. Thus, employing the GVV method, the 1st-order correction eigenstates for $|\alpha, n\rangle$ can be expanded into zeroth-order eigenstates as in the following
	\begin{equation}
		|\alpha ,n\rangle^{(1)}=\sum_{\beta,m}\frac{\langle \beta, m|\hat{\mathcal{V}}|\alpha, n\rangle }{E_{|\beta, m\rangle}-E_{|\alpha, n\rangle}}|\beta, m\rangle,\label{eq:eq6}
	\end{equation}
	with $0$th-order eigenenergy $E_{|\alpha,n\rangle} = E_{\alpha}+n\omega$, where $E_{\alpha}$ denotes energy of the physical state $|\alpha\rangle$. Then the lowest few high-order effective Hamiltonian can be computed as follows \cite{son2009floquet}:
	\begin{equation}
		\hat{\mathcal{H}}_{\mathrm{eff}}^{(1)}=\langle \Psi^{(0)}|\hat{\mathcal{V}}|\Psi^{(0)}\rangle,\label{eq:eq7}
	\end{equation}
	\begin{equation}
		\hat{\mathcal{H}}_{\mathrm{eff}}^{(2)}=\langle \Psi^{(0)}|\hat{\mathcal{V}}|\Psi^{(1)}\rangle-\hat{\mathcal{H}}_{\mathrm{eff}}^{(1)}\langle \Psi^{(0)}|\Psi^{(1)}\rangle,\label{eq:eq8}
	\end{equation}
	\begin{equation}
		\hat{\mathcal{H}}_{\mathrm{eff}}^{(3)}=\langle \Psi^{(1)}|\hat{\mathcal{V}}|\Psi^{(1)}\rangle-\hat{\mathcal{H}}_{\mathrm{eff}}^{(1)}\langle \Psi^{(1)}|\Psi^{(1)}\rangle .\label{eq:eq9}
	\end{equation}
	
	\subsection{The notations of colors}
	All colors labeled by $\Theta$ in the laser system form a set $\{\Theta| \Theta=A,B,\cdots \}$, whose subset $\{\Theta_{i}\}\subseteq \{\Theta\}$ includes colors of dressing fields on specific atom $i$. Besides, We use $\Theta^{[ij]}$ to denote the common colors between two atoms, which also form a subset $\{\Theta^{[ij]}\}=\{\Theta_{i}\}\cap\{\Theta_{j}\}$.
	
	\subsection{The full time-dependent Hamiltonian}
	Under the rotating wave approximation (RWA), the full time-dependent Hamiltonian of our system can be first written as
	\begin{equation}
		\hat{H}_{\mathrm{full}}(t)=\sum_{i}\left( \omega_{0}|r_{i}\rangle\langle r_{i}|+\sum_{\Theta}\frac{\Omega_{i\Theta}}{2}e^{-i \omega_{\Theta}t}|r_{i}\rangle\langle g_{i}|\right) +\sum_{i<j}\left(2\omega_{0}+V_{ij}\right)|r_{i}\rangle\langle r_{i}|\otimes |r_{j}\rangle\langle r_{j}|,
	\end{equation} 
	with $\omega_{0}$ the atomic transition frequency, $\Omega_{i\Theta}$ and $\omega_{\Theta}$ the Rabi frequency and laser frequency of dressing field on atom $i$ with color $\Theta$, and $V_{ij}$ the van der Waals interaction between two Rydberg atoms. By using the unitary transformation $U=\prod_{i}e^{-i\omega_{0}t|r_{i}\rangle\langle r_{i}|}$, the above time-dependent Hamiltonian will reduce to $\hat{H}_{\mathrm{full}}(t)$ in the main text.
	
	\subsection{Single-excitation}
	Though the dimension of the whole Hilbert space $L$ is large, we only need to concern a few relevant subspaces if the excitation number is conserved. For single-excitation, three subspaces should be considered: ground state $|G\rangle=| g_1 g_2 \cdots g_{\mathcal{N}}\rangle$, single-excitation subspace $\Pi_1 =\{|\Psi_{i}\rangle=| g_1 g_2 \cdots r_i \cdots g_{\mathcal{N}}\rangle, i=1,2,\cdots,\mathcal{N} \}$, and double-excitation subspace $\Pi_2=\{|\Psi_{ij}\rangle=| g_1 \cdots r_i \cdots r_j\cdots g_{\mathcal{N}}\rangle, 1\leq i<j \leq \mathcal{N} \}$. Because the dynamics we focus on occur in the single-excitation subspace, the final effective Hamiltonian can be expressed by states inside $\Pi_1 $. The Hamiltonian can be first written as
	\begin{equation}	
		\hat{H}=\sum_{i< j}V_{ij}|\Psi_{ij}\rangle \langle\Psi_{ij}|
		+\sum_{i}\left(\sum_{j\neq i,\Theta_{j}}\frac{\Omega_{j\Theta_{j}}}{2}e^{i\Delta_{\Theta_{j}}t}|\Psi_{ij}\rangle \langle\Psi_{i}|
		+\sum_{\Theta_{i}}\frac{\Omega_{i\Theta_{i}}}{2}e^{i\Delta_{\Theta_{i}}t}|\Psi_{i}\rangle\langle G|+\mathrm{H.c}\right).	\label{eq:eq10}
	\end{equation}
	Then, we assume there exist an eligible elementary frequency $\omega$ that all laser detunings $\Delta_{\Theta_{i}}$ in the Hamiltonian can be expreessed as an integer multiple, i.e. $\Delta_{\Theta_{i}}=n_{\Theta_{i}}\omega$. Thus, all Fourier components $\hat{H}_{n}$ can be found according to Eq.~(\ref{eq:eq3}). In the extended Hilbert space, the Floquet matrix can be expressed as diagonal terms $\hat{\mathcal{H}}_{0}$ and nondiagonal terms $\hat{\mathcal{V}}=\sum_{\Theta_{i}}\left(\hat{\mathcal{V}}_{-n_{\Theta_{i}}}+\hat{\mathcal{V}}_{+n_{\Theta_{i}}}\right)$, with explict forms
	\begin{equation}
		\hat{\mathcal{H}}_{0}=\sum_{n}\left( n\omega|G,n\rangle\langle G,n|+\sum_{i}n\omega |\Psi_{i},n\rangle\langle \Psi_{i},n|+ \sum_{i< j}(V_{ij}+n\omega)|\Psi_{ij},n\rangle\langle \Psi_{ij},n|\right), \label{eq:eq11}
	\end{equation}
	\begin{equation}
		\hat{\mathcal{V}}_{-n_{\Theta_{i}}}=\sum_{n}\frac{\Omega_{i\Theta_{i}}^{\ast}}{2}\left(|G,n-n_{\Theta_{i}} \rangle\langle \Psi_{i},n|+\sum_{j\neq i}|\Psi_{j},n-n_{\Theta_{i}}\rangle\langle \Psi_{ij},n|  \right), \label{eq:eq12}
	\end{equation}
	\begin{equation}
		\hat{\mathcal{V}}_{+n_{\Theta_{i}}}=\sum_{n}\frac{\Omega_{i\Theta_{i}}}{2}\left(|\Psi_{i},n+n_{\Theta_{i}} \rangle\langle G,n|+\sum_{j\neq i}|\Psi_{ij},n+n_{\Theta_{i}}\rangle\langle \Psi_{j},n|  \right). \label{eq:eq13}
	\end{equation}
	As can be seen, the operator $\hat{\mathcal{V}}_{-n_{\Theta_{i}}}$ ($\hat{\mathcal{V}}_{+n_{\Theta_{i}}}$) has two effects, one of which is deexciting (exciting) the atom on site $i$, and the other is to shift the Fourier index by $-n_{\Theta_{i}}$ ($+n_{\Theta_{i}}$). According to Eq.~(\ref{eq:eq6}), 1st-order wavefuction for $|\Psi_{i}, N\rangle$ can be calculated as
	\begin{equation}
		|\Psi_{i}, N\rangle^{(1)}=\sum_{\Theta_{i}}\frac{\Omega_{i\Theta_{i}}^{\ast}}{2\Delta_{\Theta_{i}}}|G,N-n_{\Theta_{i}}\rangle-\sum_{j,\Theta_{j}}\frac{\Omega_{j\Theta_{j}}}{2(\Delta_{\Theta_{j}}+V_{ij})}|\Psi_{ij},N+n_{\Theta_{j}}\rangle, \label{eq:eq14}
	\end{equation}
	with $N$ an arbitrary integer. Equation.~(\ref{eq:eq7})-(\ref{eq:eq9}) can then be used to calculate the effective Hamiltonian in the extended Hilbert space. It is not difficult to find that the sole nonvanishing term is $\langle \Psi^{(0)}|\hat{\mathcal{V}}|\Psi^{(1)}\rangle$ in $\hat{\mathcal{H}}_{\mathrm{eff}}^{(2)}$, which contains both diagonal and nondiagonal terms
	\begin{equation}
		\langle\Psi_{i}, N|\hat{\mathcal{H}}_{\mathrm{eff}}^{(2)}|\Psi_{i}, N\rangle=\sum_{\Theta_{i}}\frac{|\Omega_{i\Theta_{i}}|^2}{4\Delta_{\Theta_{i}}}-\sum_{j,\Theta_{j}}\frac{|\Omega_{j\Theta_{j}}|^2}{4(\Delta_{\Theta_{j}}+V_{ij})},\label{eq:eq15}
	\end{equation}
	\begin{equation}
		\langle\Psi_{j}, N-n_{\Theta_{i}}+n_{\Theta_{j}}|\hat{\mathcal{H}}_{\mathrm{eff}}^{(2)}|\Psi_{i}, N\rangle=\frac{\Omega_{i\Theta_{i}}^{\ast}\Omega_{j\Theta_{j}}}{4}\left(\frac{1}{\Delta_{\Theta_{i}}}-\frac{1}{\Delta_{\Theta_{j}}+V_{ij}} \right).\label{eq:eq16}
	\end{equation}
	Equation.~(\ref{eq:eq15}) labels the chemical potential correction to an excitation on site $i$, while hopping dynamics can be understood by further analyzing Eq.~(\ref{eq:eq16}). If there exists a \emph{channel} $\Theta^{[ij]}$ between atoms $i$ and $j$, i.e. $n_{\Theta_{i}}=n_{\Theta_{j}}\equiv n_{\Theta^{[ij]}}$, states $|\Psi_{j}, N\rangle$ and $|\Psi_{i}, N\rangle$ within the same block (labeled by $N$) can be resonantly coupled by
	\begin{equation}
		J_{ji\Theta^{[ij]}}=\frac{\Omega_{i\Theta^{[ij]}}^{\ast}\Omega_{j\Theta^{[ij]}}V_{ij}}{4\Delta_{\Theta^{[ij]}}(\Delta_{\Theta^{[ij]}}+V_{ij})}.\label{eq:eq17}
	\end{equation} 
	When such a \emph{channel} is well-defined, we can safely ignore all other nonresonant couplings between different blocks through different colors due to the large energy defect $\delta E=\Delta_{\Theta_{j}}-\Delta_{\Theta_{i}}$ between states $|\Psi_{j}, N-n_{\Theta_{i}}+n_{\Theta_{j}}\rangle$ and $|\Psi_{i}, N\rangle$. Now, the effective Hamiltonian of single-excitation subspaces within one block can be expressed as
	\begin{equation}
		\hat{\mathcal{H}}_{\mathrm{eff}}=\sum_{i}\left(\sum_{\Theta_{i}}\frac{|\Omega_{i\Theta_{i}}|^2 }{4\Delta_{\Theta_{i}}}-\sum_{j\neq i,\Theta_{j}}\frac{|\Omega_{j\Theta_{j}}|^2}{4(\Delta_{\Theta_{j}}+V_{ij})} \right)|\Psi_{i},N \rangle\langle \Psi_{i},N|+\sum_{i<j,\Theta^{[ij]}}\frac{\Omega_{i\Theta^{[ij]}}^{\ast}\Omega_{j\Theta^{[ij]}}V_{ij}}{4\Delta_{\Theta^{[ij]}}(\Delta_{\Theta^{[ij]}}+V_{ij})}|\Psi_{j},N\rangle\langle\Psi_{i},N|, \label{eq:eq18}
	\end{equation}
	which reduces to $\hat{H}_{\mathrm{eff}}^{(1)}$ in the main text.
	
	\subsection{Double-excitation}
	For the case of double-excitation, the subspaces considered include single-excitation subspace $\Pi_1$, double-excitation subspace $\Pi_2$, and triple-excitation subspace $\Pi_3$. The Hamiltonian can be expressed as
	\begin{equation}
		\begin{aligned}
			\hat{H}&=\sum_{i< j}V_{ij}|\Psi_{ij}\rangle \langle\Psi_{ij}|+\sum_{i<j<k}(V_{ij}+V_{ik}+V_{jk})|\Psi_{ijk}\rangle \langle\Psi_{ijk}|\\
			&+\left(\sum_{i}\sum_{j\neq i,\Theta_{j}}\frac{\Omega_{j\Theta_{j}}}{2}e^{i\Delta_{\Theta_{j}}t}|\Psi_{ij}\rangle \langle\Psi_{i}|+\sum_{i<j}\sum_{k\neq i,j;\Theta_{k}}\frac{\Omega_{k\Theta_{k}}}{2}e^{i\Delta_{\Theta_{k}}t}|\Psi_{ijk}\rangle \langle\Psi_{ij}|+\mathrm{H.c}\right),		
		\end{aligned}\label{eq:eq19}
	\end{equation}
	with all unperturbed diagonal terms in the first line, and the perturbed nondiagonal coupling terms in the second line. The 1st-order correction to the wavefunction $|\Psi_{ij}, N\rangle$ can be calculated as
	\begin{equation}
		|\Psi_{ij},N\rangle^{(1)}=\sum_{\Theta_{i}}\frac{\Omega_{i\Theta_{i}}^{\ast}}{2(\Delta_{\Theta_{i}}+V_{ij})}|\Psi_{j},N-n_{\Theta_{i}}\rangle+\sum_{\Theta_{j}}\frac{\Omega_{j\Theta_{j}}^{\ast}}{2(\Delta_{\Theta_{j}}+V_{ij})}|\Psi_{i},N-n_{\Theta_{j}}\rangle-\sum_{k\neq i,j;\Theta_{k}}\frac{\Omega_{k\Theta_{k}}}{2(\Delta_{\Theta_{k}}+V_{ik}+V_{jk})}|\Psi_{ijk},N+n_{\Theta_{k}}\rangle. \label{eq:eq20}
	\end{equation}
	Again, the nonvanishing diagonal and nondiagonal terms can be obtained from $\langle \Psi^{(0)}|\hat{\mathcal{V}}|\Psi^{(1)}\rangle$ in $\hat{\mathcal{H}}_{\mathrm{eff}}^{(2)}$,
	\begin{equation}
		\langle \Psi_{ij},N|\hat{\mathcal{H}}_{\mathrm{eff}}^{(2)}| \Psi_{ij},N\rangle=\sum_{\Theta_{i}}\frac{|\Omega_{i\Theta_{i}}|^2}{4(\Delta_{\Theta_{i}}+V_{ij})}+\sum_{\Theta_{j}}\frac{|\Omega_{j\Theta_{j}}|^2}{4(\Delta_{\Theta_{j}}+V_{ij})}-\sum_{k\neq i,j;\Theta_{k}}\frac{|\Omega_{k\Theta_{k}}|^2}{4(\Delta_{\Theta_{k}}+V_{ik}+V_{jk})}, \label{eq:eq21}
	\end{equation}
	\begin{equation}
		\langle \Psi_{ik},N+n_{\Theta_{j}}-n_{\Theta_{k}}|\hat{\mathcal{H}}_{\mathrm{eff}}^{(2)}| \Psi_{ij},N\rangle=\frac{\Omega_{j\Theta_{j}}^{\ast}\Omega_{k\Theta_{k}}}{4}\left(\frac{1}{\Delta_{\Theta_{j}}+V_{ij}}- \frac{1}{\Delta_{\Theta_{k}}+V_{ik}+V_{jk}}\right).\label{eq:eq22}
	\end{equation}
	
	In fact, states in $\Pi_{2}$ are not quasi-degenerate for all $ij$ due to the existence of vdW interaction $V_{ij}$. The doubly excited subspace $\Pi_{2}$ can be approximately decomposed into $\Pi_{2}=\Pi_{2}^{\prime}\cup\Pi_{2}^{\prime\prime}$, with $\Pi_{2}^{\prime}$ spanned by the dimer states and $\Pi_{2}^{\prime\prime}$ the complementary set of $\Pi_{2}^{\prime}$.
	
	\subsubsection{$\Pi_{2}^{\prime}$} 
	If the initial two excitations $i$ and $k$ are nearest-neighbors (NN), they will be bound together, forming a dimer. The dynamics for this dimer can be described as follows: when one of the excitation $i$ is viewed as fixed, its NN excitation at $k$ can hop to another NN site $j$ (i.e $V_{ij}=V_{ik}=V_{\mathrm{NN}}$) according to Eq.~(\ref{eq:eq22}) as long as there is a \emph{channel} $\Theta^{[jk]}$ connecting them.
	From Eq.~(\ref{eq:eq22}) we can work out this effective exchange interaction as $J_{jk\Theta^{[jk]}}^{(2)}\hat{b}^{\dagger}_{j}\hat{n}_{i}\hat{b}_{k}+\mathrm{H.c}$ with $\hat{n}_{i}=|r_i \rangle\langle r_i|$ and
	\begin{equation}
		J_{jk\Theta^{[jk]}}^{(2)}\approx\frac{\Omega_{j\Theta^{[jk]}}\Omega_{k\Theta^{[jk]}}^{\ast}V_{jk}}{4(\Delta_{\Theta^{[jk]}}+V_{\mathrm{NN}})(\Delta_{\Theta^{[jk]}}+V_{\mathrm{NN}}+V_{jk})},\label{eq:eq23}
	\end{equation}
	indicating that excitation $k$ can hop to site $j$ with the help of an auxiliary excitation $i$. It is noted that this dynamics is sensitive to distances between adjacent atoms because the energy defect of this hopping process is $\delta E\approx V_{ij}-V_{ik}$, thus even a few percent deviations of equal distance can destroy the quasi-degenerate condition.
	
	\subsubsection{$\Pi_{2}^{\prime\prime}$}
	Initially, if two excitations are distant enough such that $V_{ij}\ll \Delta$, excitations hop to their nearby sites with a strength following the single-excitation condition. However, the presence of a distance-dependent nonvanishing $V_{ij}$ terms in the diagonal prevent the two excitations coming too close, as if there were a repulsive interaction between excitations. In fact, this interaction further distinguishes different subspaces inside $\Pi_{2}^{\prime\prime}$. The effective Hamiltonian under this condition becomes $\hat{H}_{\mathrm{eff}}^{(2)}=\hat{H}_{\mathrm{eff}}^{(1)}+\hat{H}_{\mathrm{int}}$, with $\hat{H}_{\mathrm{eff}}^{(1)}$ the single-excitation effective Hamiltonian, and $\hat{H}_{\mathrm{int}}=\sum_{i<j}V_{ij}\hat{b}_{i}^{\dagger}\hat{b}_{j}^{\dagger}\hat{b}_{j}\hat{b}_{i}$ the nonlocal excitation-excitation interaction.
	
	\section{Tunability of Model}\label{sec:sec2}
	\begin{figure}
		\centering
		\includegraphics[width=\linewidth]{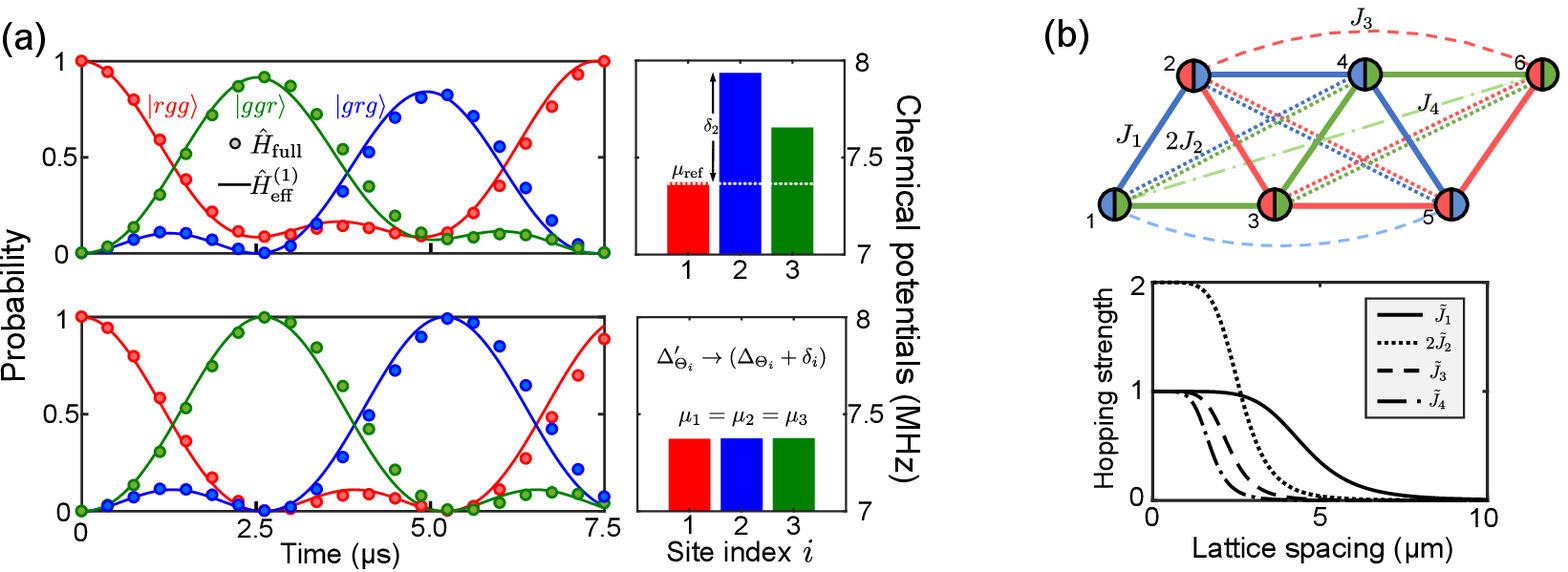}
		\caption{(a) When chemical potentials of three atoms are imbalanced, perfect chiral motion is destroyed (upper panel) but can recover after adjusting their potentials to the same reference value $\mu_{\mathrm{ref}}$ by slightly changing the detunings of some dressing fields (lower panel). (b) Different types of hopping exist when the number of sites exceeds the number of colors, whose relative strength can be tuned by changing lattice spacing $d$ while maintaining the overall shape. The normalized hopping strengths $\tilde{J_{x}}=J_{x}/J_{\mathrm{inf}} (x=1,2,3,4)$ are plotted, where $J_{\mathrm{inf}}=\Omega^2/4\Delta$ denotes a saturated value when atoms get infinitely close. }
		\label{fig:fig_s2}
	\end{figure}
	
	\subsection{Chemical Potentials}\label{sec:sec2.1}
	The chemical potential of every site can be adjusted to a more favorable value $\mu_{i}^{\prime}=\mu_{i}-\delta_{i}$ by changing the detunings of laser fields on this site slightly $\Delta_{\Theta_{i}}^{\prime}\rightarrow(\Delta_{\Theta_{i}}+\delta_{i})$, as verified in the case of three atoms [see Fig.~\ref{fig:fig_s2}(a)]. By setting $R=5~\mu m$ and keeping other parameters the same as every in the main text, chemical potentials for three sites become significantly different, leading to imperfect chiral motion for a single Rydberg excitation. After adjusting detunings of laser fields on atoms 2 and 3 by $\delta_{i}=\mu_{i}-\mu_{1}(i=2,3)$  to balance their chemical potentials to the same reference value $\mu_{\mathrm{ref}}=\mu_{1}$, the perfect chiral motion is restored. In practice, it is not as convenient to directly change the frequencies of dressing fields individually. A more feasible method could employ additional far off-resonant addressing laser beams to introduce local energy shift $\delta_{i}$ on the specific sites $i$ \cite{omran2019generation}.
	\subsection{The non-nearest neighbouring hoppings}
	Due to the limited resources available for laser frequency modulation, we should introduce as few colors as possible to assist with experimental implementation. At the same time, this consideration could cause non-nearest neighbor hopping when the number of sites exceeds the number of colors. Fortunately, the strength of non-nearest neighbor hopping relative to the nearest neighbor hopping can be tuned by adjusting distances between atoms while maintaining the same configuration. For example, if we use three colors to connect six atoms [see Fig.~\ref{fig:fig_s2}(b)], four types of hopping arise between atoms: the nearest hopping $J_{1}$ (atom $1 \leftrightarrow 2,\dots,$ etc.), the diagonal hopping $2J_{2}$ (atom $1 \leftrightarrow 4,\dots,$ etc.), the next nearest neighbour hopping $J_{3}$ (atom $2 \leftrightarrow 6,\dots,$ etc.), and the longer-range hopping $J_{4}$ (atom $1 \leftrightarrow 6$.). Their respective strengths are plotted as a function of lattice spacing $d$ in Fig.~\ref{fig:fig_s2}(b), showing that a convergent result $J_{\mathrm{int}}=\Omega^2/4\Delta$ when $d$ is small, and $J_{1}$ dominates for sufficiently large $d$.

\section{Robustness against Decoherence}\label{sec:decoherence}
In this section, we systematically study the robustness of our scheme against several major decoherence sources, including laser phase noise, thermal Doppler broadening, and decay of the Rydberg state. We show that the effective decoherence rate $\gamma_{\mathrm{eff}}$ is significantly suppressed compared with the bare decoherence rate $\gamma$ of a single Rydberg qubit, which gives rise to a favorable scaling of the error per site.

\subsection{Laser phase noise}
\begin{figure}
	\centering
	\includegraphics[width=0.8\linewidth]{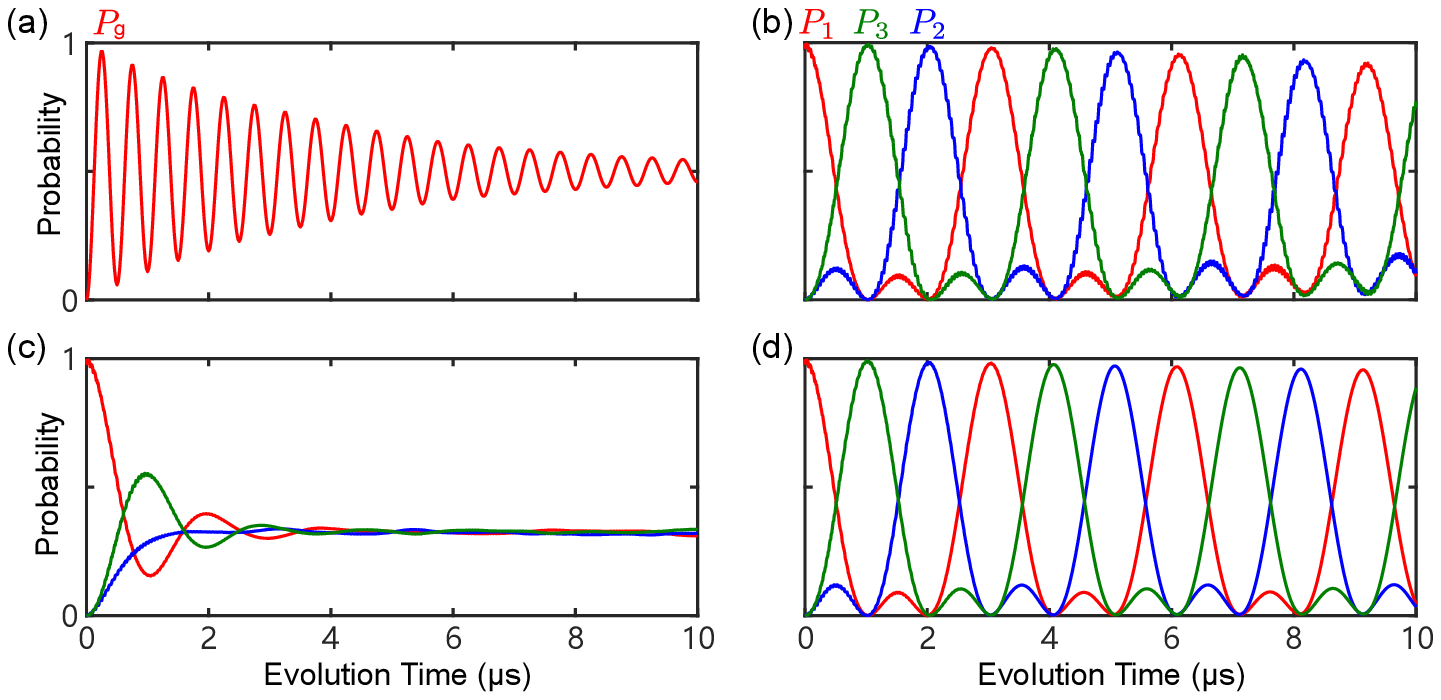}
	\caption{Laser phase noise induced decoherence calculated by $\hat{H}_{\mathrm{full}}$ at a dephasing rate $\gamma=1\mathrm{MHz}$. The phase noise is modeled by assuming a random walk of $\phi(t)$ with $\dot{\phi}(t)$ being a Gaussian white noise. (a) Single-atom damped Rabi oscillation. (b)-(d) Three-atom chiral motion under laser phase noises, which are identical for all dressing fields in (b), uncorrelated for all dressing fields in (c), and identical for the same-color fields while uncorrelated for the different-color fields in (d), respectively.}
	\label{fig:fig_PN}
\end{figure}
The laser phase noise can be described by adding a random
sequence $\phi(t)$ to the Rabi frequency as $\Omega e^{i\phi(t)}$ \cite{de2018analysis}. If all dressing fields come from the same laser source, each Rabi frequency $\Omega_{i\Theta}$ is modified by a global phase noise $\phi_{G}(t)$, and becomes $\Omega_{i\Theta} e^{i\phi_{G}(t)}$. The induced spin-exchange interaction between the $i$-th and $j$-th atom takes the
form $J_{ij\Theta^{[ij]}}\propto \Omega_{i\Theta^{[ij]}}\Omega_{j\Theta^{[ij]}}^{\ast}$, where the global phase noise $\phi_{G}(t)$ perfectly cancells out. The robustness can also be understood from the master equation description: if $\dot{\phi}_{G}(t)$ is a white noise, the evolution of the density matrix $\hat{\rho}$ can be described by the master equation $\partial_{t} \hat{\rho}=-i[\hat{H},\hat{\rho}]+\gamma\mathcal{L}[\hat{S}_{z}]\hat{\rho}$, with the Lindblad operator $\mathcal{L}[\hat{S}_{z}]\hat{\rho}=\hat{S}_{z}\hat{\rho}\hat{S}_{z}^{\dagger}-({1}/{2})(\hat{S}_{z}^{\dagger}\hat{S}_{z}\hat{\rho}+\hat{\rho}\hat{S}_{z}^{\dagger}\hat{S}_{z})$ and $\hat{S}_{z}=\sum_{i}\hat{\sigma}_{z}^{i}$, where $\hat{\sigma}_{z}^{i}=|r_{i}\rangle\langle r_{i}|-|g_{i}\rangle\langle g_{i}|$ is the Pauli $\sigma_z$ operator of the $i$-th atom. The effective Hamiltonian $\hat{H}_{\mathrm{eff}}$ has U(1) symmetry $[\hat{S}_{z},\hat{H}_{\mathrm{eff}}]=0$, which causes the dynamics to occur in a decoherence-free subspace: when simulating dynamics of $\mathcal{N}_r$ Rydberg excitations, all possible states are degenerate eigenstates of $\hat{S}_z$, which gives $\mathcal{L}[\hat{S}_z]\hat{\rho}=0$. Since the exact Hamiltonian $\hat{H}_\mathrm{full}$ does not conserve $\hat{S}_z$, we have verified the above statement by rigorous Monte Carlo simulations, i.e., modeling the noise $\phi(t)$ in the exact Hamiltonian $\hat{H}_{\mathrm{full}}$ during the three-atom chiral dynamics [corresponding to Fig.~3(a) of the main text]. As shown in Fig.~\ref{fig:fig_PN}(a), assuming a large dephasing rate $\gamma = 1 \mathrm{MHz}$, the single-qubit Rabi oscillation rapidly damps over a 10 $\mu$s timescale. Nevertheless, the chiral dynamics governed by the effective spin-exchange interaction remain coherent over much longer time scale [see Fig.~\ref{fig:fig_PN}(b)]. The inherent coherence protection relies on the correlation between phase noises from the same laser source, overcoming the otherwise rapid diffusive chiral motion if each atom experiences an independent phase noise [see Fig.~\ref{fig:fig_PN}(c)]. For our multicolor scheme, driving fields of different colors might carry independent noise due to temperature drift or stress of different fibers [see Fig.~\ref{fig:fig_exp} (b)], fields of the same color, however, always share a common phase noise. Since crosstalk between different colors is already suppressed by the secular dynamics due to the large detuning $\delta$ between different colors, the phase noise of the crosstalk is unimportant, and the dynamics remain well protected from the dephasing. We have performed calculations assuming independent phase noise for different color fields, and have confirmed that the coherence is protected [see Fig.~\ref{fig:fig_PN}(d)]. The oscillation damps even slower compared to the case of a global phase noise for all colors [Fig.~\ref{fig:fig_PN}(b)], which we attribute to the suppression of crosstalk errors (micromotions) due to the uncorrelated phase noises. With the above coherence protection feature, the final decoherence rate is determined by a small bit-flip error rate $\sim(\Omega/2\Delta)^2\gamma$ for each atom, which can be mitigated by choosing a proportionally smaller dressing parameter $\Omega/\Delta$.

\subsection{Thermal Doppler brodening}
\begin{figure}
	\centering
	\includegraphics[width=0.8\linewidth]{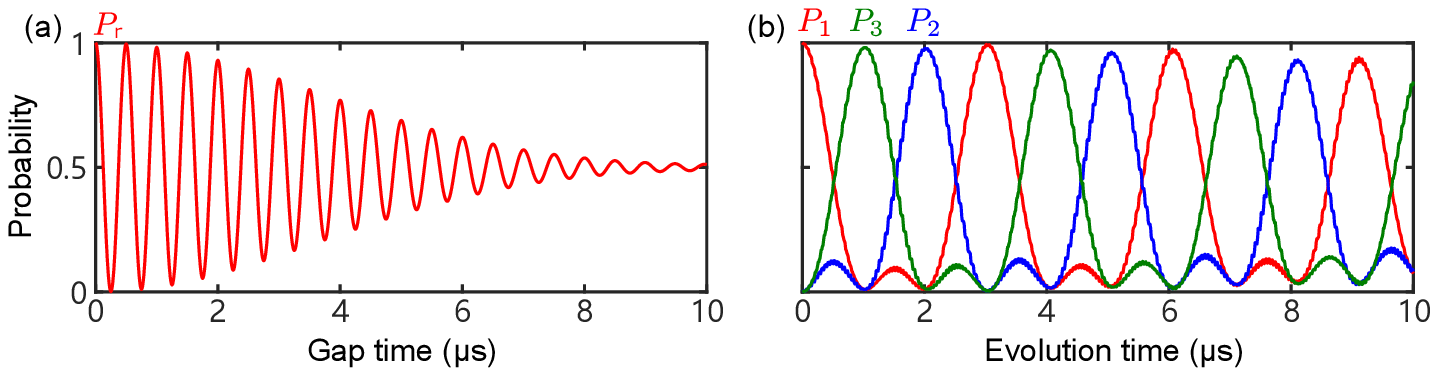}
	\caption{Thermal motion induced decoherence calculated by $\hat{H}_{\mathrm{full}}$ at a Doppler broadening of $\Delta_{T}=2\pi\times44\ \mathrm{KHz}$. (a) Ramsey fringes of a single Rydberg qubit. (b) Chiral motion of a Rydberg excitation in a three-atom plaquette. }
	\label{fig:fig_Doppler}
\end{figure}
The thermal motion of a trapped atom results in an uncertainty $\Delta_{T}$ (Doppler broadening) of the detuning for each experimental realization. For a single qubit, the Ramsey fringes rapidly damp on the time scale $1/\Delta_{T}$ [see Fig.~\ref{fig:fig_Doppler}(a) with $\Delta_{T}=2\pi\times 44\ \mathrm{KHz}$ drawn from Ref.~\cite{levine2018high}, corresponding to $T=10~\mathrm{\mu K}$]. In our scheme, such an uncorrelated dephasing creates a random on-site potential offset on the scale of $\Delta_{T}$ for the effective Hamiltonian. As a result, dephasing can be suppressed if the induced spin-exchange interaction $J$ is larger than $\Delta_{T}$. To understand this, we can first consider two-atom dynamics, where the eigenenergy of the system is given by $\pm\sqrt{J^2+ \Delta_{T}^2}\approx \pm(J+\Delta_{T}^2/J) $, i.e., the effective dephasing rate is reduced to $\Delta_{T}\times (\Delta_{T}/J)$. We have confirmed that coherence is maintained for a longer time scale by Monte Carlo simulations of the three-atom chiral dynamics in the presence of Doppler broadenings [see Fig.~\ref{fig:fig_Doppler}(b)]. For a larger system, the Doppler effect could cause localization of the Rydberg excitation, but the effective decoherence rate is suppressed as well. Assuming a simple Anderson localization, the localization length is on the scale of $\xi \sim J^2/\Delta_T^2$. According to Ref.~\cite{moix2013coherent}, the dynamics remain ballistic until the localization length is reached. Therefore, the coherence time is approximately given by $t_c=\xi/J$, which corresponds to an effective decoherence rate $\gamma_\mathrm{eff}=1/t_c\sim \Delta_T^2/J$, in agreement with the two-site analysis. On the technical side, the Doppler broadening can be easily reduced by sideband cooling of the trapped atom, as has already been demonstrated \cite{cooper2018alkaline,lorenz2021raman}.

\subsection{Rydberg decay}
\begin{figure}
	\centering
	\includegraphics[width=0.8\linewidth]{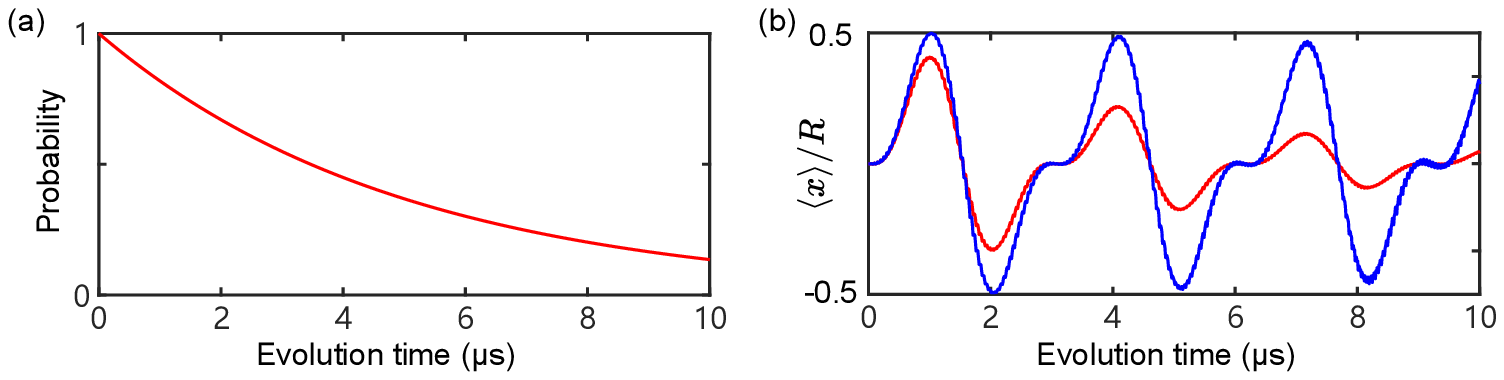}
	\caption{(a) Rapid population decay of a single Rydberg qubit at a rate $\kappa=0.2\mathrm{MHz}$. (b) Evolution of the center-of-mass coordinate $\langle x \rangle=\sum_{\nu=1,2,3}x_{\nu}P_{\nu}$ in a three-atom system with (blue line) and without (red lines) post selections at the same decay rate as in (a), where $(x_1,x_2,x_3)=(0,-R/2,R/2)$. The results are obtained by the master equation, where a time-dependent exact Hamiltonian $\hat{H}_\mathrm{full}$ is used.}
	\label{fig:fig_decay}
\end{figure}
For the Rydberg state $|70S_{1/2}\rangle$ considered in our paper, its lifetime can be up to $\tau_{r}=50~ \mu$s in existing setups \cite{levine2018high}, which is sufficiently long for the observation of the dynamics induced by the proposed effective spin-spin interactions. The decay of Rydberg excitations can be further reduced by increasing the detuning from the intermediate state in two-photon excitation schemes or using single-photon transitions, to reach $\tau_{r}=146~\mu$s at room temperature and $\tau_{r}=410~\mu$s at low temperatures by suppressing black-body radiation \cite{beterov2009quasiclassical}. In the following, we show that quantum simulations of physical observables can be purified even when Rydberg state decay remains significant. The decay of Rydberg excitations can be described by the master equation $\partial_{t}\hat{\rho}=-i[\hat{H},\hat{\rho}]+\sum_{i}\mathcal{L}[\sqrt{\kappa}\hat{\sigma}_{gr}]\hat{\rho}$, where the coherent spin dynamics are governed by the effective Hamiltonian $\hat{H}_{\mathrm{eff}}$ that conserves the total Rydberg excitation $\hat{\mathcal{N}}_{r}=\sum_{i}\hat{\sigma}_{rr}^{i}$, while the Rydberg decay causes quantum jumps which decrease $\hat{\mathcal{N}}_{r}$ by one. Therefore, if we only consider dynamics in the absence of quantum jump, evolution of the system is governed by a non-Hermitian Hamiltonian $\hat{H}_{\mathrm{NH}}=\hat{H}_{\mathrm{eff}}-(i\kappa/2)\sum_{i}\hat{\sigma}_{rg}^{i}\hat{\sigma}_{gr}^{i} = \hat{H}_{\mathrm{eff}}-(i\kappa/2)\hat{\mathcal{N}}_{r}$, where $\hat{\mathcal{N}}_{r}$ can be replaced by a number $\mathcal{N}_{r}$ because the total excitation is conserved [equivalent to the U(1) symmetry] during the no-jump dynamics, i.e., $[\hat{H}_{\mathrm{eff}},\hat{\mathcal{N}}_{r}]=0$. Therefore, $\hat{H}_{\mathrm{NH}}$ describes the same dynamics as the effective Hamiltonian $\hat{H}_{\mathrm{eff}}$ after normalization. In practice, the above purification scheme can be implemented by post-selection, which is widely practiced in quantum optics: when evaluating a physical observable in a single experimental run, only measurement outcomes that conserve $\hat{\mathcal{N}}_{r}$ are registered as a successful event. The success probability of the post-selection scales as $P=e^{-\mathcal{N}_{r}\kappa t}$, while the conditional simulation fidelity is limited by a small non-Hermitian part $\sim (\Omega/2\Delta)^2\kappa$ of the induced interaction $J$, which can be obtained by replacing $\Delta$ with $\Delta-i\kappa/2$ in its expression. The performance of the post-selection is shown in Fig.~\ref{fig:fig_decay}, where a large decay rate $\kappa=0.2\ \mathrm{MHz}$ is used. While the rapid decay of Rydberg excitations [see Fig.~\ref{fig:fig_decay}(a)] results in a proportional damping of the observable $\langle x \rangle$, the estimation of $\langle x \rangle$ based on post-selected experimental runs remains accurate for a much longer time scale [see Fig.~\ref{fig:fig_decay}(b)].

\section{Experimental Implementation}\label{sec:sec3}

	In this section, we present a scheme for experimentally implementing a flux lattice with multicolor Rydberg dressing as well as some considerations about the techniques.
	
\begin{figure}
		\centering
		\includegraphics[width=\linewidth]{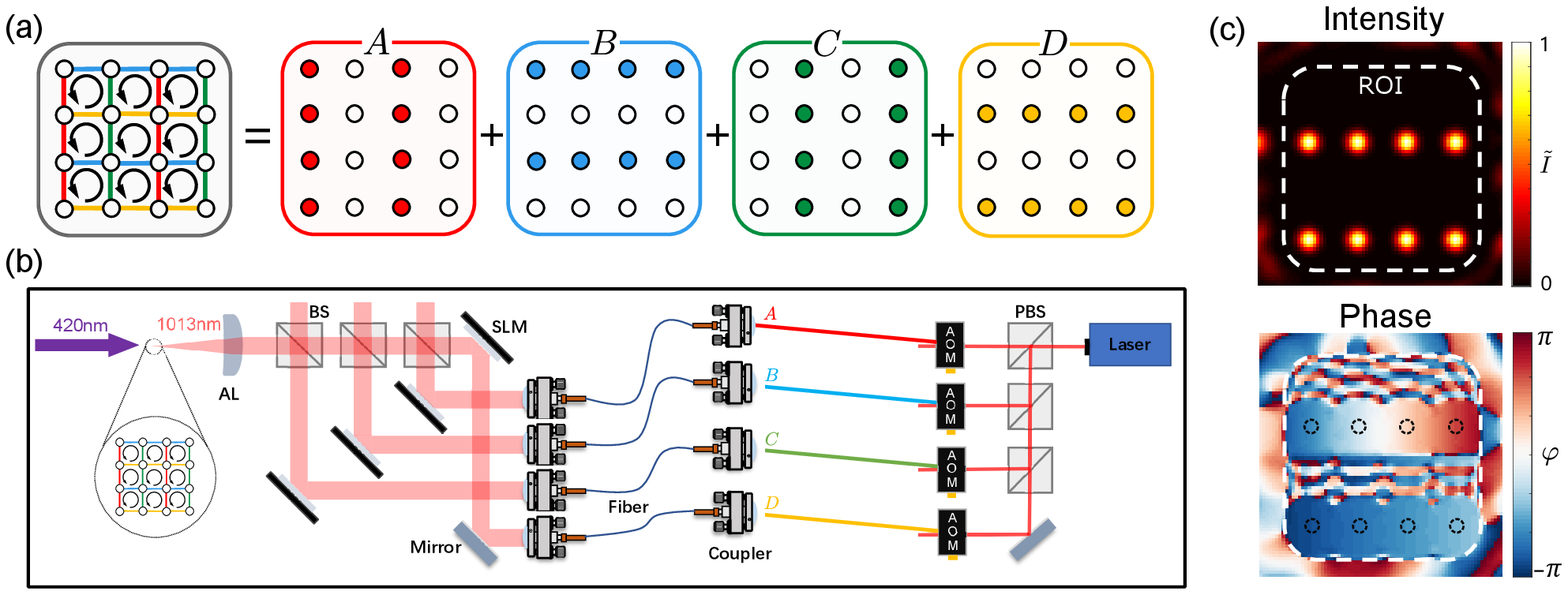}
		\caption{Schematic for implementing Harper-Hofstadter model. (a)  Four colors and their dressing patterns in a $4\times4$ square array. (b) A plausible experimental setup for two-photon dressing. The aspheric lens (AL) are used to focus addressing beams. Dressing of atoms is based on two-photon excitation scheme ($|5S_{1/2}\rangle\stackrel{420~\mathrm{nm}}{\longleftrightarrow}|6P_{3/2}\rangle\stackrel{1013~\mathrm{nm}}{\longleftrightarrow}|nS_{1/2}\rangle$) as is widely adopted. The 420~nm dressing beam illuminates atoms globally, while the addressing 1013~nm light comes with four distinguishable frequencies assuming different colors. (c) Numerical results of the 1013~nm dressing light for color $D$ on the far-field focal plane. The normalized intensity $\tilde{I}(x,y)$ and phase distribution $\varphi(x,y)$ in the region of interest (ROI) are shown respectively. }
		\label{fig:fig_exp}
	\end{figure}

	First, we show how to build up the Harper-Hofstadter lattice system considered in the main text. As illustrated in Fig.~\ref{fig:fig_exp}(b), four dressing colors $\{A,B,C,D\}$ are created by shifting the frequency of the field from a common laser source with an individual acoustic optical modulator (AOM) and are sufficient to generate individually tunable magnetic flux for arbitrarily large lattice, as one tunes the relative phases between laser fields at different sites for only two colors. This can be achieved by imprinting a computer-generated hologram (CGH) on a spatial light modulator (SLM). To obtain a proper CGH, iterative Fourier transform algorithms can be employed. However, unlike the production of a dipole trap array, for which intensity distribution is the primary concern, we need to also control the phase distribution of the laser field to construct the desired flux lattice \cite{wu2015simultaneous,bowman2017high}. To achieve such a simultaneous control of the light intensity and its phase distribution in the focal plane, we resort to the conjugate gradient minimization protocol \cite{bowman2017high}. As an example, we consider how to realize a uniform magnetic flux in the square lattice shown in Fig.~\ref{fig:fig_exp}(a), for which we choose the phases of the beam arrays in colors $B$ and $D$ to be linearly increasing along the horizontal direction, where Peierls phase corresponding to the $n$th row is $n\phi$. The generated light pattern for color $D$ is shown in Fig.~\ref{fig:fig_exp}(c), where the desired beam array can be precisely formed, with its phase well defined within the central region of each site (see the black dashed circles) and increases linearly as expected. Taking $^{87}\mathrm{Rb}$ atom as an example [see Fig.~\ref{fig:fig_exp}(b)], two-photon dressing scheme couples the ground state $|5{S}_{1/2}\rangle$ to a Rydberg state $|n{S}_{1/2}\rangle$ via the intermediate state $|6{P_{3/2}}\rangle$. Specifically, a 420~nm global laser beam couples the two lower levels ($|5S_{1/2}\rangle \leftrightarrow |6P_{3/2}\rangle$), and four 1013~nm addressing beams couple the two upper levels ($|6P_{3/2}\rangle \leftrightarrow |nS_{1/2}\rangle$). To form different \emph{channel}s, frequencies of the addressing beams are shifted correspondingly before coupling into fibers. The trapping beams (not shown in figure) can be reused to slightly adjust on-site potential offsets, the working mechanism of which has been explained in Sec.~\ref{sec:sec2.1}.	
	
	\begin{figure}
		\centering
		\includegraphics[width=0.86\linewidth]{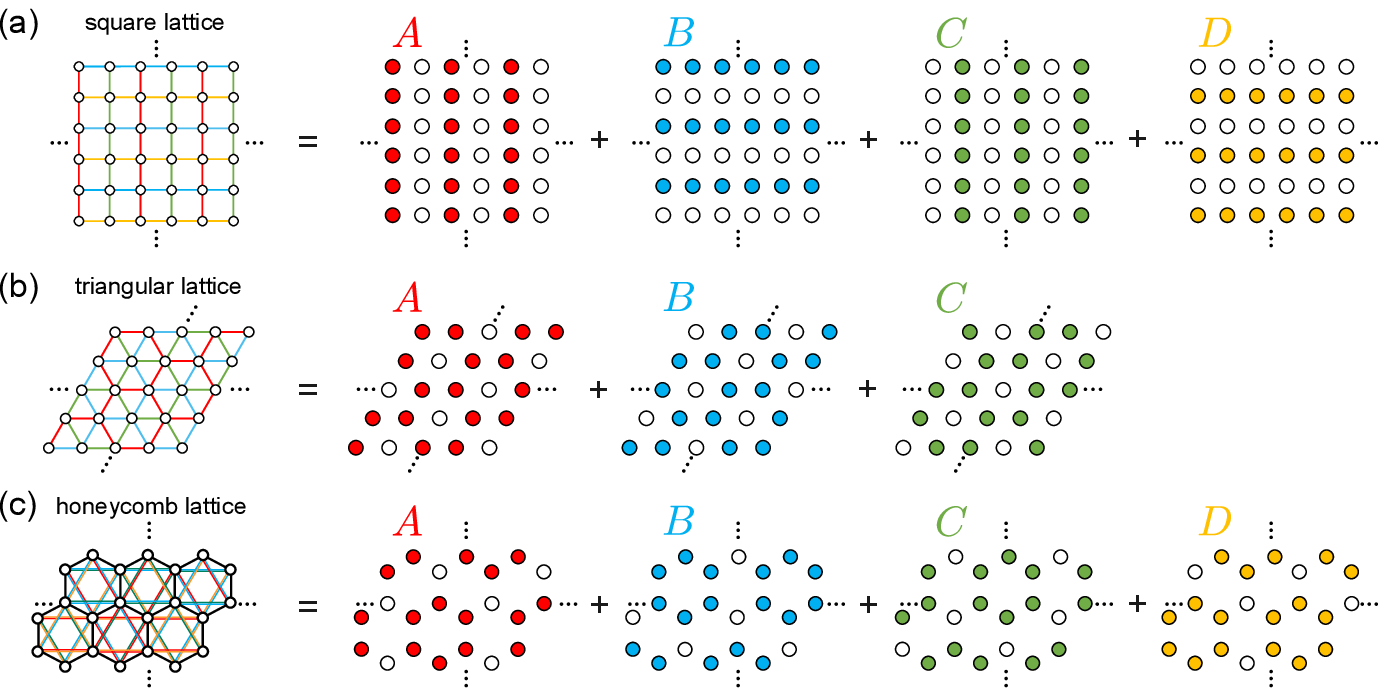}
		\caption{The number of colors does not grow with system size. (a) Dressing scheme for a square lattice with four colors. (b) Dressing scheme for a triangular lattice with three colors. (b) Dressing scheme for a honeycomb lattice with four colors.}
		\label{fig:fig_NC}
	\end{figure}
	 In the end, we discuss several technical challenges and accuracy of the effective model. First, our scheme requires single-site control of both the laser strength and phase distribution, which needs to be sufficiently homogeneous at each individual site to give a precise hopping strength $J$ and Peierls phase $\phi$. As for laser sources, the number of colors does not grow with system size. For example, to create an arbitrary magnetic flux pattern in a square lattice of an arbitrarily large size, only four different colors are required, with each site addressed by two different colors [see Fig.~\ref{fig:fig_NC}(a)], while for a triangular lattice, the number of colors is reduced to three [see Fig.~\ref{fig:fig_NC}(b)]. With four colors, it is also possible to individually manipulate the gauge flux in a Haldane honeycomb lattice [see Fig.~\ref{fig:fig_NC}(c)]. Therefore, the total power is only doubled [Figs.~\ref{fig:fig_NC}(a) and \ref{fig:fig_NC}(b)] or tripled [Fig.~\ref{fig:fig_NC}(c)] compared with conventional monochromatic Rydberg dressing schemes. The accuracy of our scheme is limited by two factors: a finite dressing parameter $\Omega/\Delta$ results in a small bit-flip error $\epsilon_b\sim(\Omega/\Delta)^2$ per site, while a finite detuning $\delta$ between different color channels incurs a crosstalk error $\epsilon_c\sim(J/\delta)^2$ between neighboring sites, where $J\sim \Omega^2/4\Delta$ denotes the effective hopping strength. For the square lattice shown in Fig.~\ref{fig:fig_exp}(a), assuming color $A$, $B$, $C$, and $D$ respectively have a detuning $(\Delta,\Delta+\delta,\Delta+2\delta,\Delta+3\delta)$, then a balanced coupling $J_A=J_B=J_C=J_D=J$ requires $\Omega_A<\Omega_B<\Omega_C<\Omega_D$. For a given strength $J$, we can obtain the laser intensity $I_{\Theta}=(\hbar/8\pi\alpha d_{\mathrm{eff}}^2)\Omega_{\Theta}^2, (\Theta=A,B,C,D)$ for each color, with $\alpha$ the Fine-Structure Constant and $d_{\mathrm{eff}}$ the effective dipole matrix element for two-photon coupling. Then, the total laser power $P_{\mathrm{tot}} \approx \pi w_{0}^2N(I_A+I_B+I_C+I_D)/2$ can be roughly estimated, where $w_{0}$ denotes the beam waist of each individual addressing field and $N$ is the size of the system. More specifically, we have
	 \begin{equation}
	 	\begin{aligned}
	 		 \Omega_{A}^2+\Omega_{B}^2+\Omega_{C}^2+\Omega_{D}^2
	 		&=\Omega_{A}^2[1+(1+\frac{\delta}{\Delta})+(1+2\frac{\delta}{\Delta})+(1+3\frac{\delta}{\Delta})]\\
	 		&=\frac{2\Omega_{A}^2}{\Delta}(2\Delta+3\delta)= 8J(2\Delta+3\delta)
	 		=8J^2(\frac{8}{\epsilon_b}+\frac{3}{\sqrt{\epsilon_c}}),
	 	\end{aligned}
	 \end{equation}
	 i.e., $P_{\mathrm{tot}}\approx  (w_{0}^2N\hbar/2\alpha d_{\mathrm{eff}}^{2})J^2(8/\epsilon_b+3/\sqrt{\epsilon_c})$. This relation shows the explicit dependency of the total laser power on the errors $\epsilon_b$ and $\epsilon_c$, and can be used to roughly estimate the laser power budget for the experiment. We also note that the design in Fig.~\ref{fig:fig_exp}(b) does not fully utilize the input laser power, and requires a few AOMs and SLMs, the number of which scales linearly with the number of colors. Improvement to reduce the overhead is also possible, e.g., by using different regions of a single SLM to modulate different color fields that are generated from a single AOD and incident obliquely upon the atom array from different directions.

	\bibliography{Supplementary}